\documentstyle[12pt,aasms4,xtex,epsfig]{article}

\def\listitem{\par \hangindent=50pt\hangafter=1
     $\ $\hbox to 20pt{\hfil $\bullet$ \hfil}}

\def\puncspace{\ifmmode\,\else{\ifcat.\C{\if.\C\else\if,\C\else\if?\C\else%
\if:\C\else\if;\C\else\if-\C\else\if)\C\else\if/\C\else\if]\C\else\if'\C%
\else\space\fi\fi\fi\fi\fi\fi\fi\fi\fi\fi}%
\else\if\empty\C\else\if\space\C\else\space\fi\fi\fi}\fi}
\def\SP{\let\\=\empty\futurelet\C\puncspace}

\def\h1{$h^{-1}$\SP}

\def\etal{{\it et al.\/}\ }

\def\eg{{\it e.g.\/}\rm,\ }
\def\lsim{~\rlap{$<$}{\lower 1.0ex\hbox{$\sim$}}}
\def\gsim{~\rlap{$>$}{\lower 1.0ex\hbox{$\sim$}}}
\def\void#1{{}}
\def\h{h^{-1}Mpc}
\def\pap1{paper~{\sc i}}
\def\pape{paper~{\sc i} }

\begin{document}

\title{Biasing and high-order statistics from the SSRS2} 
\author{C. Benoist\altaffilmark{1}}
\affil{European Southern Observatory, Karl-Schwarzschild-Str.2,
   D-85748 Garching bei M\"unchen, Germany; cbenoist@eso.org}
\author{A. Cappi}
 \affil {Osservatorio Astronomico di Bologna, via Zamboni 33, I-40126,
   Bologna, Italy; cappi@astbo3.bo.astro.it}
\author{L.N. da Costa\altaffilmark{2}}
 \affil{European Southern Observatory, Karl-Schwarzschild-Str.2,
   D-85748 Garching bei M\"unchen, Germany; ldacosta@eso.org} 
\author{S. Maurogordato\altaffilmark{1}}
\affil{Observatoire de la C\^ote d'Azur, B4229, Le Mont-Gros, F 06304,
Nice Cedex 4, France; maurogor@obs-nice.fr}
\author{F.R. Bouchet}
\affil{Institut d'Astrophysique de Paris, 98 bis Bd Arago, 75014
  Paris, France;
  bouchet@iap.fr}
 \and
\author{R. Schaeffer}
 \affil{Service de Physique Th\'eorique, CEN-Saclay, F-91191,
   Gif-sur-Yvette Cedex, France; schaeffer@amoco.saclay.cea.fr}

\altaffiltext{1}{also LAEC, CNRS, Observatoire de
Paris-Meudon, 5 Pl. J. Janssen, 92195 Meudon Cedex, France.}
 
\altaffiltext{2}{also Departamento de Astronomia CNPq/Observat\'orio
Nacional, rua General Jos\'{e} Cristino 77, Rio de Janeiro, R.J. 20921
Brazil.} 

\begin{abstract} 

We analyze different volume--limited samples extracted from the
Southern Sky Redshift Survey (SSRS2), using counts--in--cells to
compute the Count Probability Distribution Function (CPDF).From the
CPDF we derive volume--averaged correlation functions to fourth order
and the normalized skewness and kurtosis $S_3 =
\bar{\xi_3}/\bar{\xi_2}^2$ and $S_4=\bar{\xi_4}/\bar{\xi_2}^3$. We
find that the data satisfies the hierarchical relations in the range
$0.3 \lsim \bar{\xi_2} \lsim 10$. In this range, we find $S_3$ to be
scale-independent with a value of $\sim 1.8$, in good agreement with
the values measured from other optical redshift surveys probing
different volumes, but significantly smaller than that inferred from
the APM angular catalog. In addition, the measured values of $S_3$ do
not show a significant dependence on the luminosity of the galaxies
considered. This result is supported by several tests of systematic
errors that could affect our measures and estimates of the cosmic
variance determined from mock catalogs extracted from {\it N-body}
simulations. This result is in marked contrast to what would be
expected from the strong dependence of the two-point correlation
function on luminosity in the framework of a linear biasing model. We
discuss the implications of our results and compare them to some
recent models of the galaxy distribution which address the problem of
bias.

\end{abstract}


\section{Introduction}

One of the simplest measures of clustering is the two-point
correlation function. However, it provides a complete description only
in the case of a Gaussian distribution.  The complexity of the
large-scale clustering pattern of galaxies (large voids, great walls,
groups, clusters) reveals non-Gaussian features that cannot be fully
described by second order statistics. Instead, a distribution of $N$
objects can only be fully described by its series of $N$-point
correlation functions or alternatively, by the entire set of count
probabilities $P(N,V)$, i.e. the probability of finding $N$ objects in
a randomly placed volume $V$. This set defines the count probability
distribution function, hereafter CPDF.

Direct calculation of the two, three and four-point 
correlation functions on galaxy catalogs has shown that the few-body 
correlations on
mildly non-linear scales (i.e., from $\sim 1$ to a few $\h$) can 
be expressed as a sum of products of two-point correlation functions:
\begin{equation}
\xi_3(r_1,r_2,r_3) = Q [\xi_2 (r_1) \xi_2(r_2) + \xi_2 (r_2) \xi_2(r_3) +
\xi_2 (r_1) \xi_2(r_3)],
\label{trepunti}
\end{equation}
with values of $Q$ in the range 0.8 to 1.3 (\eg Groth \& Peebles 1977,
Fry \& Seldner 1982, Efstathiou \& Jedredjewski 1984).
This relation has been generalized to the $N$-point correlation function
of the matter distribution, defining 
a set of hierarchical models (Fry 1984a, Schaeffer 1984): 
\begin{equation}
\label{hier}
\xi_J ({\mathbf r}_1,...,{\mathbf r}_J) =
\sum_\alpha Q_J ^{(\alpha)} \sum_{ij} \prod^{J-1} \xi_2(r_{ij}),
\end{equation}
where the $Q_J^{(\alpha)}$ are independent of the scale and $\alpha$
denotes each irreducible geometrical configuration of $J$-points.
 In 
general $Q_J$ depends on the shape of the $J$-tuple 
$({\mathbf r}_1,...,{\mathbf r}_J)$, although the dependence in the 
case of $Q_3$ should be fairly weak (Fry 1984b).
These hierarchical models can also be characterized by the
volume-averaged correlation functions $\bar{\xi}_J = V^{-J} \int_V
\xi_J d^3 r_1 ...d^3 r_J$ through the relation:
\begin{equation}
\bar{\xi}_J(r) = S_J \bar{\xi_2}^{J-1}(r),
\label{averhier}
\end{equation}
where the coefficients $S_J$ are independent of the scale and are 
related to the previously defined $Q_J$ by (Balian \& Schaeffer 1989):

\begin{equation}
S_J = B_J J^{J-2}Q_J.
\end{equation}
The $B_J$ are geometrical factors close to 1, which depend on
the shape of $\xi(s)$. 

The hierarchical relation (\ref{hier}) is a solution of the BBGKY
equations in the strongly non-linear regime and is also predicted in
the linear regime, as shown by using perturbation theory (\eg
Juszkiewicz, Bouchet \& Colombi 1993). However, the value of $Q_J$
($S_J$) may differ in the different regimes. Therefore, an important
question is whether this hierarchical relation is satisfied by the
mass distribution and on what range of scales it is valid. This also
raises the problem of how to relate these predictions for the mass
fluctuations to the observed galaxy distribution, namely the nature of
the biasing process.

In order to further investigate these issues we analyze the clustering
properties of galaxies using the Southern Sky Redshift Survey (SSRS2,
da Costa \etal 1994, 1998). Its depth, dense sampling and geometry
make it one of the most suitable samples currently available to study
high order statistics over a wide range of scales.  In previous papers
(Benoist \etal 1996, hereafter \pap1, Cappi \etal 1998), we have used
the SSRS2 catalog to investigate the dependence of galaxy clustering
on luminosity and the nature of galaxy bias using second order
statistics.  In the present paper we extend our work to higher order
investigating the behavior of the skewness $S_3$ and the kurtosis
$S_4$ as functions of luminosity. In particular, we address the
following questions: 1) the validity of the hierarchical relations; 2)
the range of scales over which these relations are satisfied ; 3) the
nature of bias based on the properties of higher order moments, more
specifically the skewness.

These issues have been addressed in different papers.  For instance,
Bouchet \etal (1993) have examined the validity of the hierarchical
relations using the 1.2 Jy IRAS redshift survey. However, this sample
under-represents high density regions and does not adequately probe
nonlinear clustering, because of the low density of galaxies. The use
of the skewness and kurtosis to probe the nature of biasing was
investigated by Gazta\~naga and Frieman (1994), using the
two-dimensional APM galaxy survey.

In section 2 we briefly describe the data and the various sub-samples
used in the present analysis. We also present the CPDF derived from
the SSRS2 and compare them to various analytical models that have been
proposed to describe the observed clustering properties of
galaxies. In section 3, we compute the volume averaged correlation
functions up to the fourth order and examine the validity of the
hierarchical model.  We also derive the skewness and the kurtosis for
the various sub-samples. In section 4, we investigate the dependence
of the skewness on the luminosity, examining possible sources of
errors in its determination. In section 5. these results are compared
to the predictions of different galaxy biasing models.  Finally, in
section 6 we summarize our main conclusions.

\section{Counts-in-Cells}

In the present analysis we use the SSRS2 south sample (da Costa \etal
1994, 1998) which contains about 3600 galaxies brighter than
$m_B=15.5$, distributed over 1.13 steradians of the southern galactic
cap ($b<-40^{o}$) within the declination range
$-40^{o}<\delta<-2.5^\circ$.  Radial velocities have been corrected to
the Local Group rest frame (Yahil, Tammann, \& Sandage 1977). For
comparison we also use the SSRS2 north (da Costa \etal 1998) and CfA2
south samples (Huchra \etal 1997), covering smaller areas ($\sim 0.6$
steradians) in the northern and southern galactic caps, respectively.
Even though smaller, these samples are useful to directly gauge the
variance of the various statistics.

Since all galaxies in the SSRS2 have been assigned morphological
types, following the ESO-Uppsala classification system ($T$), we have
adopted a morphological-dependent K-correction, $K(z,T)$, to compute
the absolute magnitude, $M$, of a galaxy according to 
\begin{equation}
M = m - 25 -5 \log D_L - K(z,T),
\end{equation}
where $D_L$ is the luminosity distance for standard Friedman models:
\begin{equation}
D_L = \frac{c}{H_0} \frac{1}{q_0^2}
    [1 - q_0 + q_0 z + (q_0 -1) \sqrt{2 q_0 z +1}].
\end{equation}
 In the following analysis, we have adopted $H_0 = 100$ and $q_0 =
0.5$.  We have divided the sample into four morphological classes and
applied for each class the K-correction following Efstathiou, Ellis \&
Peterson (1988) (see \pap1).  The analysis presented below has been
carried out in redshift space.

\placetable{tab_id}

We have extracted from the SSRS2 six volume-limited sub-samples as
done in \pap1. The characteristics of these sub-samples are given in
Table~\ref{tab_id} which lists: in column (1) the sub-samples id; in
column (2) the depth; in column (3) the absolute magnitude limit; in
column (4) the number of galaxies; in column (5) the density; in
column (6) the correlation length $s_0$.  Note that the magnitude
interval between sub-samples is not a constant because of the strong
luminosity segregation observed in \pap1 for galaxies brighter than
$L_*$, which requires a finer grid at the bright end.

We have measured $P(N,s)$ as the fraction of randomly placed spheres
of radius $s$ containing exactly $N$ galaxies.  The probabilities are
computed using $10^6$ random spheres (in order to have less than
$10\%$ error on $P(N,s=1h^{-1}Mpc)$), rejecting those crossing the
survey edges.  While this procedure limits the number of independent
spheres on large scales, it has the advantage of making no assumptions
about the galaxy distribution beyond the surveyed volume.

In Figures \ref{fig_pn195}-- \ref{fig_pn21} we show the CPDF computed
for the sub-samples D91, D138, and D168 respectively, compared to the
predictions of four models: the gravitational quasi-equilibrium
(Saslaw \& Hamilton 1984, Saslaw \etal 1990, Itoh, Inagaki \& Saslaw
1993, Sheth, Mo \& Saslaw 1994); the negative binomial (Carruthers \&
Minh 1983); the log-normal (Coles \& Jones 1991); and the Gaussian
distribution.

\placefigure{fig_pn195}

\placefigure{fig_pn205}

\placefigure{fig_pn21}

On small scales ($s \lsim 10\h$), where significant deviations from
Gaussianity are noticeable, all analytical models fit the data equally
well over the whole range of $N$, independently of the luminosity
class.  On larger scales, there is a clear trend towards Gaussianity
as examplified in panels (b), (c) and (d) of Figure \ref{fig_pn205},
corresponding to scales in the range $20\h \lsim s \lsim 30\h$. Note
that in this case all analytical models provide a poor fit for
intermediate values of $N$. Unfortunately, several effects plague this
comparison and prevent the use of the CPDF to discriminate between
models.  For fainter samples (figure \ref{fig_pn195}) the volume
probed is too small to reach Gaussianity. In other cases, the samples
are too dilute to either compute reliable counts or to show
significant differences between models. Despite these problems it is
still true that using the CPDF can be, in some cases, more informative
than quantities related to the moments of the distribution. Since the
moments are an integral quantity they have larger errors. It is worth
pointing out that the CPDF computed even for the
extreme case of sub-sample D168, containing the brightest
galaxies, which has the smallest number of galaxies and lowest
density, looks comparable to the other more dense sub-samples.

\section {Moments of the Counts Distribution}

\subsection{Relation between Correlation Functions}

The volume-averaged correlation functions have been calculated from
the counts in cells.  The centered moments of order $J$, $\mu_J$, can
be directly derived from the count probabilities $P(N,s)$ (Peebles
1980):

\begin{equation}
\mu_J(s) = \langle (N - \bar{N})^J\rangle
 = \sum_{N=0}^{N_{max}(s)} P(N,s) (N - \bar{N})^J,
\end{equation}
where $N_{max}(s)$ is the maximum number of galaxies counted in a
sphere of radius $s$, and $\bar{N}=\sum NP(N)$, i.e. at each scale we
have computed directly the effective mean number of galaxies from the
counts, and not from simple ratio $N_{gal}/V_{sample}$.
 
In the case of a smoothed distribution, the volume-averaged
correlation functions  
$\bar{\xi_J} = V^{-J} \int_V \xi_J d^3 r_1 ...d^3 r_J$ are related to the 
corresponding cumulants through $\bar{N}^J\bar{\xi}_J=k_J$, where
$k_2=\mu_2$, $k_3 = \mu_3$, $k_4 = \mu_4-3\mu_2^2$
(Kendal \& Stuart 1991).
%
For a discrete distribution, these relations have to be corrected for
shot-noise. We assume that our galaxy samples are Poisson
realizations of an underlying smooth ``galaxy'' density field.  Thus,
we obtain the following relations between the volume-averaged correlation
functions and the cumulants of the discrete distribution (Peebles
1980, Fry 1985):
\begin{eqnarray}
\bar{\xi}_2&=&
\frac{1}{\bar{N}^2}(k_2-\bar{N})\nonumber\\
\bar{\xi}_3&=&
\frac{1}{\bar{N}^3}(k_3-3k_2+2\bar{N}) \nonumber\\
\bar{\xi}_4&=&
\frac{1}{\bar{N}^4}(k_4-6k_3+11k_2-6\bar{N})
\label{xibar} 
\end{eqnarray}

Following this procedure, we have computed the volume-averaged
correlation functions up to order 4 for the six volume-limited
sub-samples presented in Table~\ref{tab_id}. $\bar{\xi}_2$ is shown in
Figure \ref{fig_ksi}, where the change in the amplitude as a function
of the luminosity can be easily seen.  Here, as well as in the
subsequent plots, the error bars correspond to the rms value as
measured for 15 bootstrap re-samplings of each sub-sample. We have
measured the increase in the amplitude of $\bar{\xi}_2(s)$ in the
range $2\h \lsim s \lsim 10\h$ and derived the relative bias ratios
for the different luminosity classes, assuming a linear biasing
model. As discussed in \pape the relative bias shows a weak dependence
on luminosity for sub-L$_*$ galaxies, with a rapid increase at the
bright end.

\placefigure{fig_ksi}

In order to estimate the amplitude of sample-to-sample variations, in
Figure \ref{fig_ksicomp} we compare $\bar{\xi}_2(s)$ for each SSRS2
south sub-sample to those obtained from similar sub-samples extracted
from the CfA2 south and SSRS2 north. Note that, for D168 this
comparison was not possible because of the small number of bright
galaxies in these two catalogs which cover about half of the area
surveyed by the SSRS2 south. As it can be seen, within 10$\h$ there is
a remarkable agreement in the amplitude and shape of the
$\bar{\xi}_2(s)$ derived from these surveys probing different
directions of the sky and different large-scale structures.  Most of
the differences occur on larger scales, where the results are
sensitive to cosmic variance.

\placefigure{fig_ksicomp}

For each sub-sample we have also computed the volume-averaged
three-point $\bar{\xi}_3$, and four-point $\bar{\xi}_4$, correlation
functions. In Figure \ref{fig_hr} these functions are shown as a
function of $\bar{\xi}_2$ for the six sub-samples considered.  Also
shown in each panel is the best power-law fit to the data points
corresponding to $L_*$ galaxies (D91). As predicted by the
hierarchical model there is a tight relation between the third and
fourth order correlations with $\bar{\xi}_2$. Moreover, the slopes of
the power-law fits are in good agreement with the expected index
$(J-1)$ in the framework of the hierarchical models, as defined by
equation (3). For comparison, in Figure \ref{fig_hr} we also show the
best fit given by Bouchet \etal (1993) for the IRAS galaxies.  The
comparison between IRAS and SSRS2 galaxies shows that these two
populations follow similar relations with comparable amplitudes.

\placefigure{fig_hr}

It is worth pointing out that the hierarchical relations hold to
relatively large values of the variance ($\bar{\xi}_2 \sim 10$),
corresponding to non-linear scales.  It is important to emphasize that
the optical data used here gives further support to the original claim
of Bouchet \etal (1993) that the hierarchical relation extends to
relative non-linear regime, since with the IRAS data scales
corresponding to $\bar{\xi}_2 \sim 10$ could only be probed with a
single sub-sample. While this behavior is predicted by perturbation
theory for the mildly nonlinear regime, there is currently no
compelling theoretical reason to justify this behavior for the
non-linear scales probed by the optical data, except that the solution
for the BBGKY equations has this property.  Some deviations from the
scale-invariance are noticeable on small ($\bar{\xi}_2 \lsim 0.2$) and
large scales ($\bar{\xi}_2 > 10$). A key question is to determine
whether these deviations are due to the limitations of the present
data or reflect a real break in scale-invariance.

The above results are summarized in Table~\ref{tab_hr} which gives: in
column (1) the sub-sample id; in columns (2) and (3) the slope of
$\bar{\xi}_3-\bar{\xi_2}$ and $\bar{\xi}_4-\bar{\xi}_2$, respectively;
in column (4) the limit of $\bar{\xi_2}$ below which the relations
deviate from the hierarchical model, and in column (5) the related
scale range.

\placetable{tab_hr}

\subsection {Skewness and Kurtosis}

Another way of expressing the hierarchical relations is to examine the
behavior of the skewness and kurtosis as a function of scale. This has
been used to compare with different cosmological models and to
constrain the galaxy biasing process (\eg Gazta\~naga \& Frieman
1994).

In Figure~\ref{fig_sn} we show the behavior of $S_3$ and $S_4$ as a
function of scale, as derived from the SSRS2 south. From the upper
panel we find that, in general, $S_3$ is roughly constant for all
samples, over a range scales. For some samples deviations can be seen
on small ($\lsim 3\h$) and large scales ($\gsim 10\h$). Note that the
values of $S_3$ for the different samples are comparable, yielding
$\langle S_3 \rangle \sim 1.8$.  The reasons for the deviations from
the scale-invariant behavior are discussed in more detail below. It is
worth pointing out that apart from these deviations, our results
strongly suggest that dependence of $S_3$ on luminosity is very
weak. Taking it a face value this is an important result, because it
is in marked contrast with the expectations of linear
biasing. However, several effects may compromise these measurements,
as discussed in the next section.

\placefigure{fig_sn}

In the bottom panel of Figure~\ref{fig_sn}, we also show the computed
values for $S_4$. In contrast, to the skewness it shows strong
fluctuations as a function of scale and large errors. This is not
surprising since higher order moments are increasingly more sensitive
to various effects related to the limitations of the
data. Nevertheless, each sample seems to be characterized by a value
of $S_4$ which ranges from about 3 to 10. Another interesting point is
that brighter samples tend to yield the largest values of $S_4$, also
in contradiction with the predictions of a simple linear bias model.
However, as in previous case a firm conclusion can only be drawn after
understanding the limitations of the data. Because of the extreme
sensitivity of $S_4$, in the rest of the paper we focus our attention
primarily on the skewness.

\subsection{Dependence of $S_3$ on Luminosity}

One of the primary goals of the present paper is to investigate the
nature of galaxy biasing using higher order moments such as the
skewness, which for our samples are significantly more robust than the
kurtosis (Gazta\~naga \& Frieman 1994). Using the results of the
previous section we compute for each volume-limited sub-sample an
average value of $S_3$. This is done over the range of scales where
the scale-invariance is valid.  The results are shown in Figure
\ref{fig_bias}, where we plot $S_3$ as a function of the absolute
magnitude difference relative to a $L_*$ galaxy. The error bars
associated to the data points were computed using bootstrap.

\placefigure{fig_bias}

In the framework of linear biasing, where $\delta_{g}(\mathbf x ) =
b_g\delta(\mathbf x )$, one can easily predict the dependence of
high-order moments on the luminosity of the galaxies. In this case,
the averaged $J$-point correlation function for galaxies is related to
mass $J$-point correlation function by:
\begin{equation}
\bar{\xi}_{J,g}= b_g^J \bar{\xi}_J,
\end{equation}
which implies that the quantities $S_J$ (equation \ref{averhier}), should
vary as 
\begin{equation}
S_{J,g} = \displaystyle{\frac{S_{J}}{ b_g^{J-2}}}.
\label{snbias}
\end{equation}

Assuming that in general the biasing factor may depend on the
luminosity, we can define a relative bias, $b_g/b_{g^*}$,
normalized to its  measured value at some characteristic luminosity, such
as $L_*$, in which case one gets:
\begin{equation}
\left(\frac{S_{J,g}}{S_{J,g^*}}\right)=\left(\frac{b_g^*}{b_{g}}\right)^{J-2}
\,\,\,\,\,\,\,\,\,
{\rm for} \,\,\,\, J>2.
\end{equation}

From the above is easy to see that in the case of linear bias the
skewness should depend inversely on the biasing factor. This, of
course, presumes that one may define a unique value for these higher
order moments as a function of luminosity. To compare these
predictions for the linear biasing model with those measured for the
SSRS2, we show in Figure \ref{fig_bias} the relative variation
expected for $S_3$, given the relative bias derived in \pape from the
variance $\bar{\xi}_2$. In that paper, it was shown that galaxies
brighter than $L_*$ are more clustered than fainter galaxies, based on
the dependence of the variance, determined on scales $\lsim 10
\h$. However, the results based on second order analysis cannot by
themselves constrain the biasing model, which is the main motivation
for considering the behavior of the skewness.

\placetable{tab_s3}

The quantitative results of this comparison are summarized in
Table~\ref{tab_s3} which lists: in column (1) the sub-sample id;
in columns (2) and (3) the values of $S_3$ computed from the data and
expected from the linear bias model; in column (4) the separation
interval used in the calculation of the average values of $S_3$ from
the data. We point out that the errors given in the table include the
uncertainties in the estimates of $b_g/b_{g^*}$ (\pap1).

As anticipated, there is a clear disagreement between the measured
values and those predicted by the linear biasing model which predicts
larger values of $S_3$ for faint galaxies and smaller values for
brighter galaxies than observed. From the data we find that that $S_3$
is roughly constant, with a mean value of $\sim 1.8$ and a dispersion
of $\sim 0.1$ relative to the mean.  Furthermore, based on the
statistical errors alone these differences are significant
($\gsim~3~\sigma$) for the samples D50, D74, and D168.  Even though
not shown in the figure this trend remains true even for shallower
sub-samples, such as that containing galaxies brighter than M~=~-17,
taking into account the cosmic variance as measured from mock samples
extracted from the N-body simulations (see below).

It is interesting to note that similar values of $S_3$ have been found
for clusters of galaxies even though they are several times more
biased than galaxies (Cappi \& Maurogordato 1995; see also Plionis \&
Valdarnini 1995; Gazta\~naga, Croft, \& Dalton 1995), which again
would argue against a simple linear biasing model.

Our results tend to favor non-linear bias of the galaxy distribution
on scales $\lsim 10\h$, the range used in estimating $S_3$ from the
data. However, even though suggestive, these results rely on: 1) that
the effects which might influence our estimate of the skewness (see
discussion below) are luminosity independent; and 2) that the
bootstrap errors are a fair representation of the cosmic variance in
our measurements.

\section {Discussion}

\subsection {Amplitude of $S_3$}

From the six SSRS2 sub-samples, we find that the average values $S_3 =
1.8 \pm 0.1$ and $S_4 = 5.5\pm 1.0$, over the range of scales where
these quantities are roughly constant.  As mentioned above, these
values are higher than those derived from the 1.2 Jy IRAS redshift
Survey by Bouchet \etal (1992, 1993), who obtained $S_3=1.5\pm 0.5$,
and $S_4=4.4\pm 3.7$. On the other hand, our measurements are in good
agreement with those derived by Gazta\~naga (1992) from shallower
optical redshift surveys such as CfA1 ($S_3 = 1.86 \pm 0.07$, $S_4 =
4.15 \pm 0.6$) and SSRS ($S_3 = 2.01 \pm 0.13$, $S_4 = 4.96 \pm
0.88$). Similar results have also been obtained by Ghigna \etal (1997)
who have used the Pisces-Perseus Redshift Survey (Giovanelli \& Haynes
1989), finding $S_3 \sim 2.2$, $S_4 \sim 6.5$. The good agreement
between these various results is interesting because these surveys
probe different directions and volumes, sampling independent
structures.

On the other hand, as shown in previous works, the skewness and
kurtosis as derived from nearby redshift surveys are significantly
smaller than those derived from the analysis of the projected
distribution of APM Galaxy Survey ($S_3=3.16\pm 0.14$,
$S_4=20.6\pm2.6$; Gazta\~naga 1994). A similar result has been found
in the case of IR galaxies with the projected IRAS galaxy catalog
($S_3=2.19\pm 0.18$; Meiksin, Szapudi \& Szalay 1992).

The discrepancy between the values obtained from two-dimensional and
three-dimensional analysis may have different causes not all of them
well understood. Among them are: 1) finite statistics effects, due to
the small density of galaxies in some sub-samples; 2) redshift
distortions; 3) sampling effects, due to the small volumes probed by
the redshift surveys, which may cause one to miss rare high
multiplicity events; and 4) cosmic variance. These effects plague not
only the amplitude of the high-order moments but may also affect the
way these moments depend on the scale of the measurement.  On the
other hand, analysis of two-dimensional samples may be affected by
projection effects and by mixing galaxies over a wide range of
luminosities (Ghigna \etal 1997).

The effect of analyzing low-density samples have been investigated by
Colombi, Bouchet \& Schaeffer (1994,1995; hereafter, CBS1 and
CBS2). In principle, equations (\ref{xibar}) are reliable only in the
regime where the mean number of particles $\bar{N}$ is large. To
investigate the impact of this effect directly from the data, we have
diluted our 6 sub-samples, by randomly extracting a smaller number of
galaxies in each sub-sample, and analysed the impact on the moments of
CPDF in the small $\bar{N}$ regime.

\placefigure{fig_dilu}

In Figure \ref{fig_dilu}, we illustrate the low-density effect on
$\bar{\xi}_2$ and $S_3$ by diluting the sub-sample D91. In the figure
we show the original sample, and subsets containing a factor of 2 and
4 less galaxies. For each dilution factor, eight subsets were used and
are represented by their mean values, with the error-bars representing
the rms deviation.  We find that while $\bar{\xi}_2$ is not
significantly affected, $S_3$ is sensitive to the density of the
sample. The main effect is to decrease the amplitude of $S_3$,
primarily on small scales, changing the shape of $S_3$ as a function
of scale, which in extreme cases does not even show a well defined
plateau.  Note that the low-density has little effect on large-scales.
The above probably explains why in Figure~\ref{fig_dilu} we see a
gentle decrease on small scales of the $S_3$ value measured for the
subsamples D138 and D168, which are the least dense subsamples
considered in the present analysis.

The impact of redshift distortions on the skewness and kurtosis is
still a matter of debate. Some authors have argued that this effect
may account for the observed differences between the values for $S_3$
derived from two- and three-dimensional samples (Bouchet \& Hernquist
1992, Lahav \etal 1993, Matsubara \& Suto 1994).  More recently,
Bonometto \etal (1995) and Ghigna \etal (1997) have shown, for several
CHDM and CDM models, that this is not necessarily the case, in
agreement with empirical findings of Fry \& Gazta\~naga (1994), using
CfA1 and SSRS. It is worth pointing out that as shown by Willmer, da
Costa \& Pellegrini (1998), redshift distortions in SSRS2 south are
small, in contrast to other regions of the sky.

The estimate of $S_3$ based on the moments of the CPDF is also
sensitive to size of the sample, as the inclusion or exclusion of rare,
high-multiplicity systems affect the large--$N$ tail of the
distribution which in turn impacts the high order moments. As shown by
Kim \& Strauss (1997) this may bias the value derived, for instance,
for $S_3$, especially on large scales. This may explain the
differences between the values derived from the APM projected
distribution and from redshift surveys, as mentioned above. However,
we are primarily interested in investigating the possible impact on
the relative amplitudes of $S_3$ as a function of luminosity and not
on the absolute value of the skewness.

In order to investigate this effect, we have applied to our optical
data the method recently proposed by Kim \& Strauss (1998), whereby
the values of $\bar{\xi}_2$ and $S_3$ are determined as parameters in
a maximum likelihood fit of an Edgeworth expansion to third order (\eg
Juszkiewicz \etal 1995), convolved with a Poissonian distribution to
the observed CPDF. The method assumes an ad hoc shape for the CPDF and
can only be applied for scales in the mildly non-linear regime and for
which several independent volumes are available. Therefore, it cannot
replace the curves obtained for $S_3$ from the moments method.
Instead, we have selected, for each sub-sample, scales that satisfy the
conditions of validity of this approximation.  The results of the fits
are shown in figure \ref{fig_edge} for four volume--limited samples of
the SSRS2, at a scale where our estimate of the variance $\bar{\xi}_2$
from the moments method is between $0.8$ and $0.9$. The values for the
variance and skewness are given in Table \ref{tab_edge}, where we
list: in column (1) the sample id; in column (2) the radius of the
sphere; in columns (3) and (4) the values of $\bar{\xi}_2$ and $S_{3}$
computed directly with the moments method; in columns (5) and (6) the
values of $\bar{\xi}_{2E}$ and $S_{3E}$ obtained with the Edgeworth
model.

\placefigure{fig_edge}

\placetable{tab_edge}

While the moments method gives values
of $S_3$ around 1.5, the Edgeworth model gives larger values ($S_3
\sim 3$) in good agreement with the findings of Kim \& Strauss
(1998). However, we stress that in our case the value of $S_3$ was
computed where scale--invariance is observed 
(corresponding to $\bar{\xi}_2 > 1$), 
while the Edgeworth method requires $\bar{\xi}_2 < 1$.
It is therefore not trivial to compare the absolute values found 
by both methods. 
More important for our purpose is that the amplitude of
$S_3$ does not depend on the magnitude limit of the sample, in
agreement with the results obtained from the moments method.

The analysis of Kim \& Strauss (1997) also gives a natural explanation
for the observed decrease of $S_3$ (Figure \ref{fig_sn}) on large
scales. In particular, their figure 3, shows the larger sensitivity of
$S_3$ on large scales to the tail of the CPDF. Using the Edgeworth
method they also show that scale-invariance extends to very large
scales, which justifies assigning a single value of $S_3$ for each
volume-limited sub-sample.

\placefigure{fig_s3_ksi}

\subsection{Measurement Errors in  $S_3$}

Another crucial point in the interpretation of our results is to
appropriately estimate the errors associated with our measurements of
$S_3$ for the different sub-samples. Until now the errors shown have
been estimated using the bootstrap method which does not take into
account the cosmic variance. In order to evaluate the latter we have
used both the data themselves and mock catalogs extracted from
$N$-body simulations.

One measure of the cosmic variance can be obtained directly by
comparing the measurement of $S_3$ for the different sub-samples as
determined from the SSRS2 south, north and CfA2 south, as shown in
figure \ref{fig_s3comp}. In general, there is good agreement on
intermediate scales. However, strong deviations can be seen on large
scales due to the smaller volumes probed by CfA2 south and SSRS2
north. Note that for $L_*$ galaxies (D91) the agreement is excellent
between SSRS2 south and CfA2 south.

\placefigure{fig_s3comp}

Even though valuable to gauge the importance of cosmic variance, the
comparison between different catalogs does not provide us the means to
assign errors to our measurements.  To do that we have resorted to
mock catalogs extracted from $N$-body simulations using the same
survey geometry and selection function as the SSRS2 south (Borgani,
private communication). The simulations have a box size of 250$\h$,
include $128^3$ particles, and are normalized to COBE.  Galaxies were
identified with the particles of the simulation which were selected to
produce the same number density as the SSRS2.  Four different CDM
models have been used : standard CDM, open CDM ($\Omega_0=0.6$;
$\Omega_{\Lambda}=0$; $\Omega_{b}=0.024/h^2$), $\Lambda$ CDM
($\Omega_0=0.4$; $\Omega_{\Lambda}=0.6$; $\Omega_{b}=0.024/h^2$), and
tilted CDM ($\Omega_0=1$; $\Omega_{b}=0.05/h^2$).  In each case,
twelve independent mock SSRS2 samples have been extracted, and have
been used to compute more realistic estimates of the cosmic variance.

$S_3$ has been derived using the same procedure as for the real
catalog, for the same limiting magnitude as in figure \ref{fig_sn}. We
present the results related to the standard CDM (Figure \ref{fig_cdm})
and to the tilted CDM (Figure \ref{fig_tcdm}), which represent two
extreme cases in terms of cosmic variance.  The data points represent
the mean value obtained for twelve realizations and the error bars the
standard deviation. Comparing these to the errors estimated from the
bootstrap method we find that the latter tends to underestimate the
true error by a factor $\lsim 2$ in the case of the standard CDM.
This can be seen in Figure~\ref{fig_bias} where the errors derived
from the mock catalogs are also displayed.  In this case, the
differences between the linear bias predictions and the data are still
significant. The case of a tilted CDM makes it difficult to conclude
for the brightest galaxies, but the conclusion remains valid for the
subsamples corresponding to the faintest galaxies.

\placefigure{fig_cdm}

\placefigure{fig_tcdm}

\section{Comparison with Theoretical Models}

Given the failure of the simple linear bias model to explain the
absence of any significant dependence of higher order moments on
luminosity in our analysis of the SSRS2, we investigate the
possibility that the observed variation can be explained by non-linear
effects on the biasing process.  Such possibility has been examined by
other authors trying to explain the differences between the skewness
of optical and infrared galaxies (FG93, Juszkiewicz \etal 1995) or
between optical galaxies and some of the most popular cosmological
models (Gazta\~{n}aga \& Frieman 1994).

Here we investigate whether a non-linear model, where $\delta_g =
f(\delta)$, could account for the observed dependence of the skewness
on luminosity.  In the linear or quasi-linear regime ($\delta \sim
1$), we can expand $f$ in a Taylor series (FG93):
\begin{equation}
\delta_g=f(\delta)=b_{0,g}+\sum_{k=1}^{\infty}\frac{b_{k,g}}{k!}\delta_{mass}^k,
\label{bnonlin}
\end{equation}
where $b_{1,g,}$ is the usual linear bias parameter $b$, and $b_{0,g}$
is calculated in order to have $\langle \delta_g\rangle=0$.  The
skewness of the galaxy distribution is then related to that of the matter
distribution by the equation:
\begin{eqnarray}
S_{3,g}&=&b^{-1}\left[S_{3,mass}+3c_{2,g}\right]+O(\bar{\xi}_2),
\label{s3nonlin}
\end{eqnarray}
with $c_{k,g}=b_{k,g}/b$. In the following, we will always consider
two classes of galaxies ``$g$'' and ``$g^*$'' (corresponding to $L_*$
galaxies, characterized by a biasing factor $b^*$). Therefore, for
sake of clarity, we will hereafter omit the subscripts ``g'', with the
understanding that each quantity depends on the class of objects
considered. The above equation can be rewritten to relate the skewness
between  two different classes of galaxies, eliminating the
unobservable skewness of the mass distribution leading to (FG93)
\begin{equation}
S_3=C_1^{-1}\left[S_3^*+3C_2\right],
\label{s3relnonlin}
\end{equation}
where the coefficients $C_k$ define the biasing transformation between
the classes of galaxies $g^*$ and $g$, or equivalently between
$\{b^*;b_2^*;b_3^*...\}$ and $\{b;b_2;b_3...\}$, and $S_3^*$
corresponds to the value of the skewness for the D91 sub-sample.  The
first two terms are given by
\begin{eqnarray}
C_1&=&\frac{b}{b^*}\nonumber\\
C_2&=&\frac{b_2/b-b_2^*/b^*}{b^*}.
\label{ck}
\end{eqnarray}

In Table \ref{tab_ck} we present for the six sub-samples (column 1),
the values of $C_1$ (calculated in \pap1) and $C_2$ (estimated from
equation \ref{s3relnonlin}) in columns (2) and (3), respectively.  The
fact that $C_1\ne1$ means that galaxies from different luminosity
classes are biased relatively to each other, while non-zero values
of $C_2$ measure the deviation from linear biasing.  This result is also
presented in Figure~\ref{fig_ck}, where the relative second order
bias, $C_2$, as a function of magnitude is compared to several models
described below.

\placetable{tab_ck}

\placefigure{fig_ck}

Mo, Jing \& White (1996) (hereafter MJW) have developed an analytical
model which makes specific predictions for $S_J$, and therefore for
the coefficients $C_1$ and $C_2$ which can be computed from the
data. This was done for dark matter halos in the quasi-linear regime,
by extending the Press-Schechter formalism to include dynamical
effects (see also Mo \& White 1996).  In order to compare our results
to this model, we make the crude assumption of a one-to-one relation
between halos and galaxies.

While the general relations derived by MJW do not allow a direct
comparison with the data, these authors show that the expressions for
the coefficients $b_k$ can be simplified considering several
asymptotic cases. The first assumption we use is the fact that halos
should be identified at low redshifts. Then, the two following cases
have been investigated : big halos, leading to $b_k\approx b^k$, and
small halos, leading to $b_k=k!(a_{k-1}+a_k)(1-b)$ (where the $a_k$
are coefficients appearing in the expansion related to the spherical
collapse model, see MJW). Therefore, in the case of big halos
identified at low redshift we obtain :
\begin{equation}
C_2 = C_1-1, 
\label{eqck}
\end{equation}
\\
and in 
the case of small halos identified at low redshift :
\begin{equation}
C_2 = \frac{8}{21}\frac{1}{b^*}\left(1-\frac{1}{C_1}\right). 
\label{eqck2}
\end{equation}

The model also predicts $b/b_*$ as a function of mass (Mo \& White
1996). As seen in \pap1, the case of a low bias ($\sigma_8 \sim 1$)
provided the best representation of the SSRS2 data at the faint-end,
but it did not match the rapid rise of the relative bias for brighter
galaxies. This shortcoming of the model reflects the lack of
understanding of the galaxy-halo connection.  In Figure~\ref{fig_ck},
we compare $C_2$, calculated from SSRS2 data, to the $C_2$ predicted
by MJW for both big halos and small halos, at low redshifts. In these
comparisons we have used the values of $C_1$ estimated from the SSRS2
(see \pap1). Therefore, we can directly test the relation between
$C_1$ and $C_2$, independently from any prediction for $C_1$. Note
that to make the comparison in the case of small halos, we need to
provide the value of $b^*$ which is taken to be $b^* = 1$.

Despite the crude assumption of identifying halos to galaxies, and the
fact that we have only considered asymptotic cases of the model
proposed by MJW, it is remarkable that the resulting trend is quite
similar to the one observed from the data.  Notice that the case of
small halos provides a better fit to the data for galaxies fainter
than $L_*$, and big halos show a steep increase of $C_2$ similar to
the brightest galaxies of the SSRS2.  These two asymptotic cases bound
the measured values of $C_2$ as a function of luminosity. A more
quantitative comparison would require a better understanding of the
galaxy-halo connection.

An alternative approach is to describe the non-linear density field at
the present epoch, assuming that biasing is produced only by
gravitational processes, and that the $N$-body matter correlation
functions obey the hierarchical model (Schaeffer 1984, 1985). Like in
the previous case this model is able to predict the relative bias as a
function of luminosity as well as the relations between high order
moments (Bernardeau \& Schaeffer 1992,1996). Qualitatively, the model 
predicts a very weak
variation with the luminosity, with typically $S_3 \sim 2.5$, 
consistent with our result.
However, we also point
out that, as in the case of the Mo \& White model, the predicted
behavior of $(b/b^*)$ as a function of luminosity shows some
deviations as compared to the data, as shown in \pap1.

\section{Conclusion}

Using the SSRS2 sample we have computed counts-in-cells from which
volume-averaged correlation functions and related skewness and
kurtosis have been derived. These quantities have been used to test
the validity of the hierarchical model and to investigate the nature
of galaxy bias.  Our main conclusions can be summarized as follows :
\begin{itemize}
\item[1)]
The CPDF analysis discriminates only weakly the
various models, although it has shown a tendency towards Gaussianity
on scales $\sim 25\h$, that would need larger and denser samples to be
confirmed. 
\item[2)]
We have confirmed the hierarchical relations,
$\bar{\xi}_J=S_J\bar{\xi}_2^{J-1}$,  for $0.3\lsim \bar{\xi}_2\lsim 10$,
and $J=3,4$.
\item[3)]
We have computed the skewness over a wide range
of magnitudes (from $M=-18$, up to $M=-21$), and have found them to
show only a very weak
dependence on luminosity over the scale range from $\sim 2$ to $\sim
10\h$. 
\item[4)]
We have shown that the previous result is in
contradiction with the linear biasing scheme, $\delta_g=b_g\delta_{mass}$. 

\item[5)] The weak variation of $S_3$ with luminosity can be explained
by a non-linear biasing model. In particular, we make specific
predictions for the dependence of the relative second order bias on
luminosity.

\item[6)] Even though none of the current theoretical models for
biasing can reproduce our results in detail, all show a weak
dependence of the skewness on luminosity, as seen in the data.

\end{itemize}

Wide-angle nearby surveys such as SSRS2 and CfA2 offer an adequate
geometry and a sufficiently large number of galaxies for the analysis
of counts-in-cells for different volume-limited samples. From this
analysis the high-order clustering properties and their dependence on
luminosity can be investigated. These surveys represent the best
currently available three-dimensional samples for this kind of
analysis.  While our results strongly favor a non-linear biasing
process, confirmation will have to await ongoing deep wide angle
surveys such as 2dF or SDSS.

\bigskip
\acknowledgments

We would like to thank all the SSRS2 collaborators. LNdC would like to
thank the hospitality of the Institut d'Astrophysique de Paris and the
Observatoire de Meudon where part of this work was carried out. We
thank Stefano Borgani for providing us his CDM mock catalogs and for
helpful discussions. We are also grateful to Adi Nusser and Simon
White for fruitful discussions. The authors wish also to thank the
anonymous referee for her/his comments.


\newpage

\begin{center}
\begin{deluxetable}{cccccc}
\tablecaption{Sub-Samples of the SSRS2. \label{tab_id}}
\tablewidth{0pt}
\tablehead{
\colhead{Sub-sample}& \colhead{D} & \colhead{$M_{lim}$} & \colhead{ N$_{g}$} & 
\colhead{$\bar{n}$} &\colhead{$s_0$}
\nl
 \colhead{}&\colhead{($\h$)} & \colhead{} &\colhead{}    
& \colhead{($10^{-3}h^3$Mpc$^{-3}$)} & \colhead{($\h$)}  }

\startdata
D50 &50.12 &-18.0 &487 &11.8  &3.5$\pm$0.2
\nl
D74 &73.86 &-19.0 &770 &5.1   &5.5$\pm$0.4
\nl
D91 &91.37 &-19.5 &755 &2.6   &5.8$\pm$0.6
\nl
D112 &112.60 &-20.0 &593 &1.1 &5.5$\pm$0.6
\nl
D138 &138.12 &-20.5 &268 &0.3 &8.0$\pm$0.4
\nl
D168 &168.52 &-21.0 &133 &0.1 &15.8$\pm$2.9
\enddata
\end{deluxetable}
\end{center}

\newpage

\begin{center}
\begin{deluxetable}{ccccc}
\tablecaption{$\bar{\xi}_2-\bar{\xi}_3$ and $\bar{\xi}_2-\bar{\xi}_4$ \label{tab_hr}}
\tablehead{
\colhead{Sub-sample} 
&
\colhead{$\displaystyle{\left(\frac{{\rm d}\log\bar{\xi}_3}
{{\rm d}\log\bar{\xi}_2}\right)_{\bar{\xi}_2< 
\bar{\xi}_{2,lim}}}$}  
&\colhead{$\displaystyle{\left(\frac{{\rm d}\log\bar{\xi}_4}
{{\rm d}\log\bar{\xi}_2}\right)_{\bar{\xi}_2<\bar{\xi}_{2,lim}}}$} 
&\colhead{$\bar{\xi}_{2,lim}$} &\colhead{Scale range ($\h$)}}
\startdata
D50  &2.0$\pm$0.04  &3.1$\pm$0.2  &0.8 &2-5
\nl
D74  &2.1$\pm$0.07  &3.5$\pm$0.3  &0.7 &2-8
\nl
D91  &2.1$\pm$0.05  &2.9$\pm$0.2  &1.0 &2-7 
\nl
D112 &1.9$\pm$0.05  &2.7$\pm$0.2  &0.8 &2-8
\nl
D138 &2.0$\pm$0.08  &3.0$\pm$0.2  &0.6 &2-11
\nl
D168 &2.1$\pm$0.09  &3.1$\pm$0.2  &0.3 &2-17
\enddata
\end{deluxetable}
\end{center}

\newpage

\begin{center}
\begin{deluxetable}{cccc}
\tablecaption{$S_3$ \label{tab_s3}}
\tablehead{
\colhead{Sub-ample} & \colhead{$S_3$} & \colhead{$S_3^*b^*/b$} 
& \colhead{Interval} 
\nl
\colhead{} &\colhead{}& &\colhead{($\h$)}}
\startdata
D50  &2.0$\pm$0.12 &3.1$\pm$0.3  &1.2-5.0 
\nl
D74  &1.8$\pm$0.14 &2.4$\pm$0.2  &1.2-6.0 
\nl
D91  &2.2$\pm$0.11 &2.2$\pm$0.2  &1.2-6.0
\nl
D112 &1.6$\pm$0.14 &2.2$\pm$0.2  &2.5-10.0 
\nl
D138 &1.7$\pm$0.12 &1.5$\pm$0.1  &5.0-9.0 
\nl
D168 &1.9$\pm$0.15 &1.0$\pm$0.1  &3.0-30.0 
\enddata
\end{deluxetable}
\end{center}

\newpage

\begin{center}
\begin{deluxetable}{ccrrrr}
\tablecaption{Estimates of $\bar{\xi}_2$, and $S_3$ \label{tab_edge}}
\tablehead{
\colhead{Sub-sample} &\colhead{Scale} &\colhead{$\bar{\xi}_2$}
&\colhead{$S_3$} &\colhead{$\bar{\xi}_{2E}$} &\colhead{$S_{3E}$}
\nl
\colhead{} &\colhead{($\h$)} &\colhead{} &\colhead{} &\colhead{} &\colhead{}}
\startdata
  D50 &  5.0  & 0.87 $\pm$ 0.11 & 1.58 $\pm$ 0.16  & $0.57 \pm 0.10$ & $3.08 \pm 0.48$ 
\nl
  D74 &  7.0  & 0.90 $\pm$ 0.13 & 1.70 $\pm$ 0.18 & $0.66 \pm 0.24$ & $2.93 \pm 0.49$ 
\nl
  D112 & 7.9  & 0.85 $\pm$ 0.12 & 1.57 $\pm$ 0.20 & $0.56 \pm 0.03$ & $3.01 \pm 0.25$ 
\nl
  D138 & 11.0 & 0.81 $\pm$ 0.15 & 1.67 $\pm$ 0.23 & $0.63 \pm 0.15$ & $2.84 \pm 0.19$ 
\enddata
\end{deluxetable}
\end{center}

\newpage

\begin{center}
\begin{deluxetable}{ccc}
\tablecaption{$C_1$ and $C_2$ the first two parameters of
  the biasing transformation between the referenced sub-sample D91 and
  the other sub-samples. \label{tab_ck}}
\tablewidth{0pt}
\tablehead{
\colhead{Sub-sample}& \colhead{$C_1$} & \colhead{$C_2$}  }

\startdata
D50 &0.71$\pm$0.03 &-0.25$\pm$0.10
\nl
D74 &0.91$\pm$0.05 &-0.18$\pm$0.10
\nl
D91 &1.00$\pm$0.04 &0.00$\pm$0.10 
\nl
D112 &1.02$\pm$0.06 &-0.19$\pm$0.12 
\nl
D138 &1.44$\pm$0.09 &0.09$\pm$0.12  
\nl
D168 &1.87$\pm$0.20 &0.55$\pm$0.22 
\enddata
\end{deluxetable}
\end{center}

\newpage

\figcaption[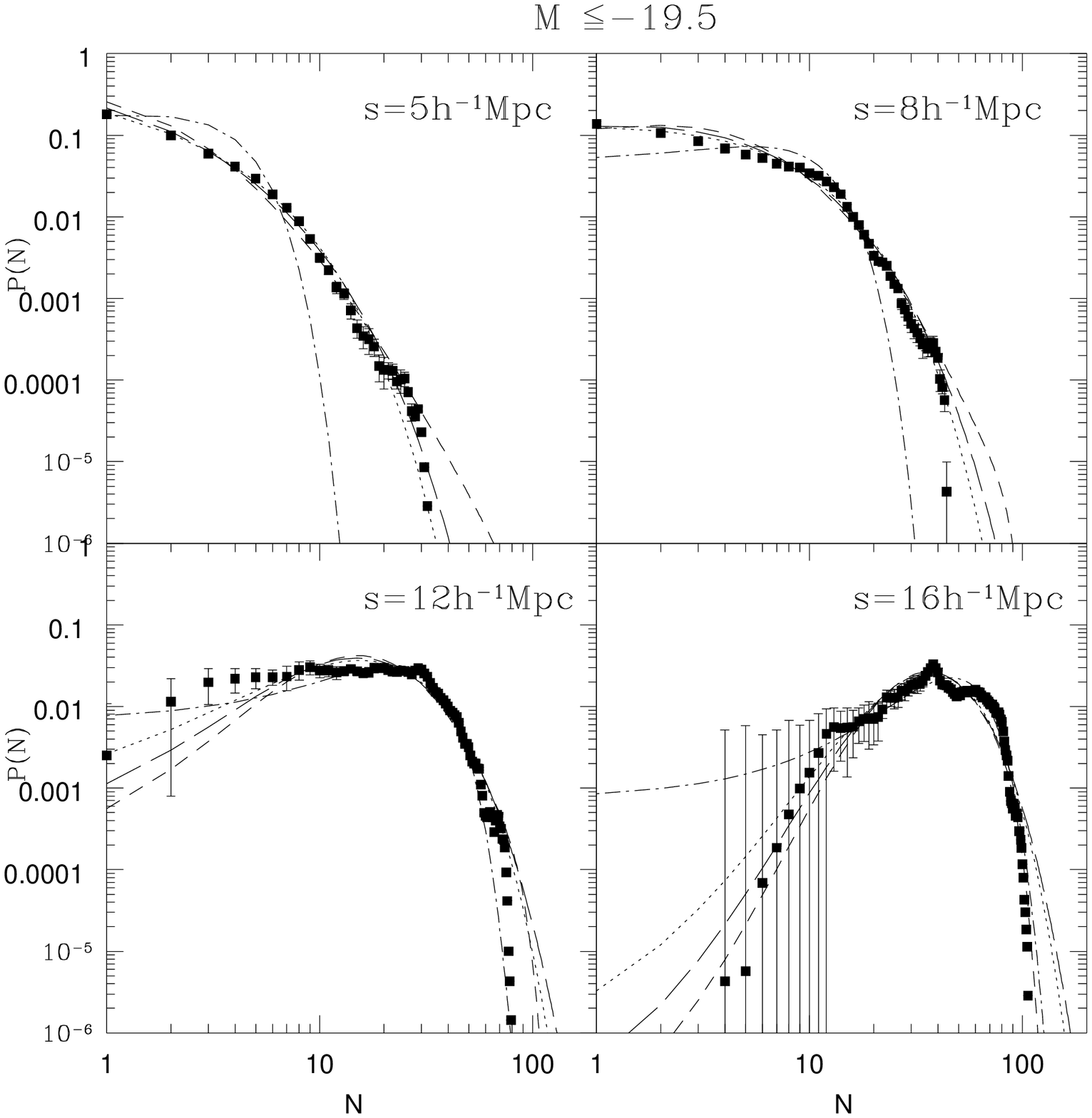]{$P(N)$, for
the D91 sample (dots), compared to the following theoretical models:
the negative binomial model (dotted line), the lognormal model (Coles \& Jones)
(dashed line), Saslaw's model(long-dashed line), the Gaussian model
(dashed-dotted line).}

\figcaption[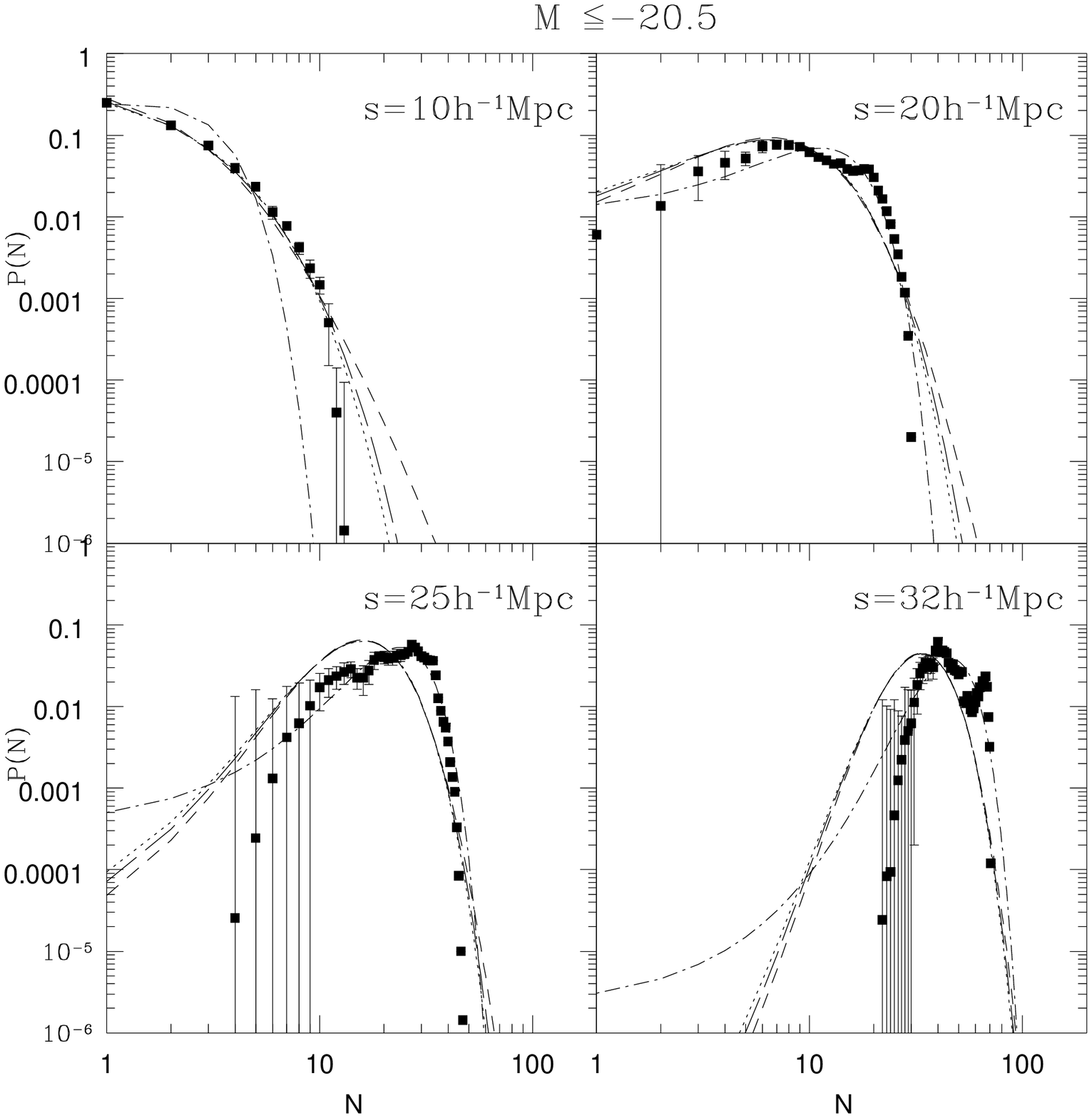]{$P(N)$, for
the D138 sample (dots), compared to the same theoretical models as in
Figure \ref{fig_pn195}.}

\figcaption[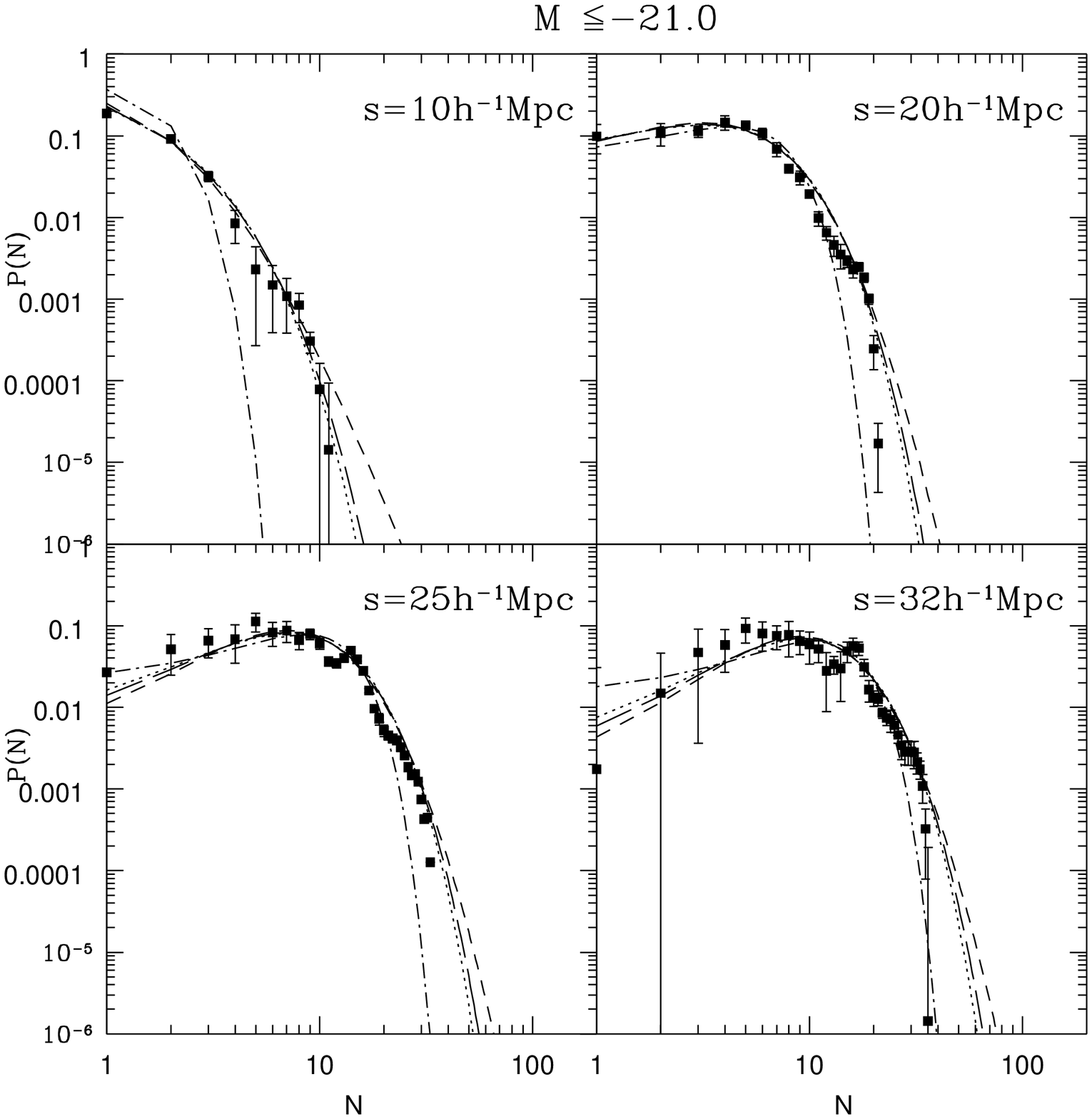]{$P(N)$, for
the D168 sample (dots), compared to the same theoretical models as in
Figure \ref{fig_pn195}.}

\figcaption[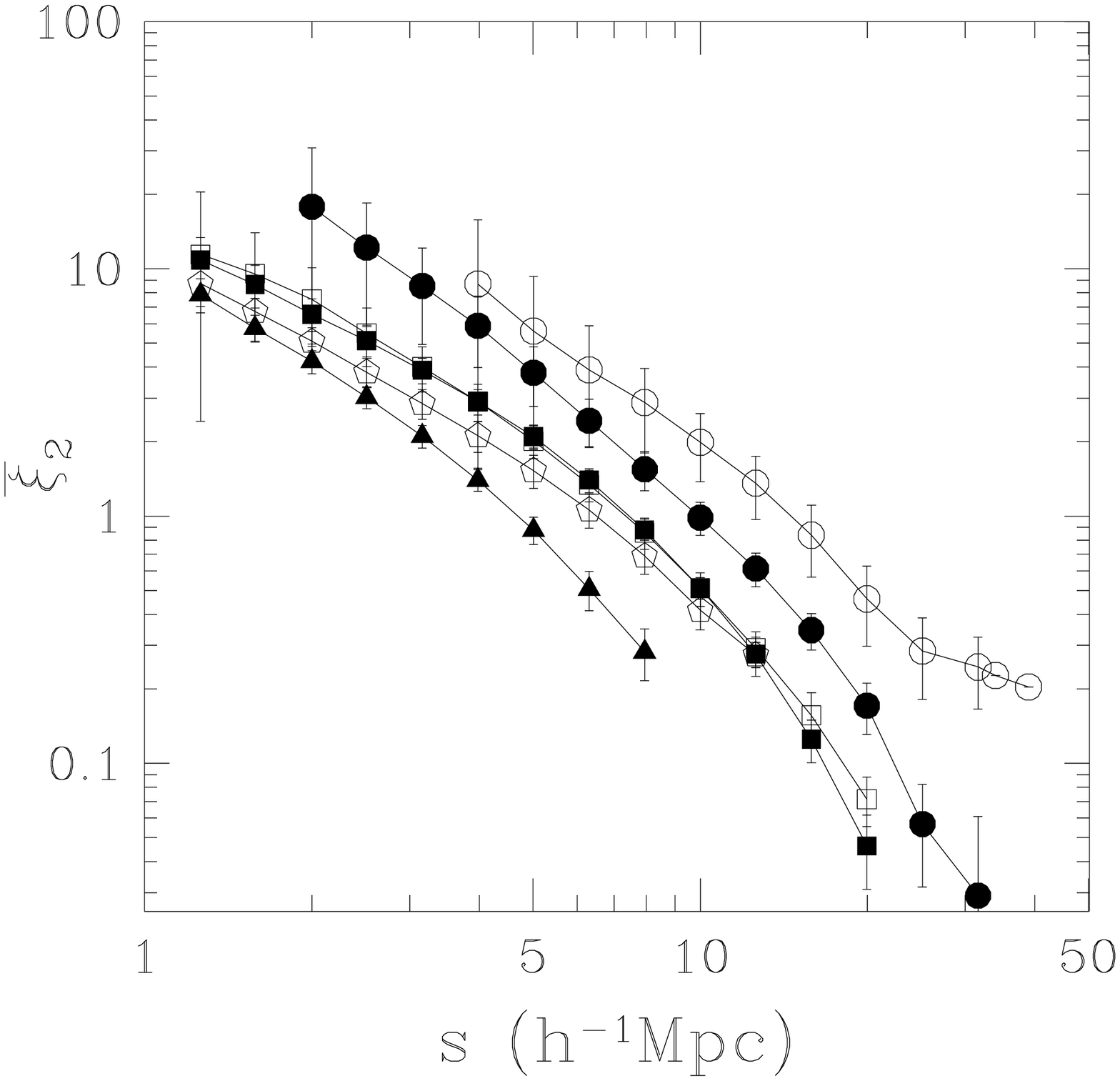]{The average two-point correlation functions 
for various volume-limited sub-samples identified as follows : 
D50 (full triangles); D74 
(open pentagons); D91 (full squares); D112 (open
squares); D138 (full circles); D168 (open circles).}

\figcaption[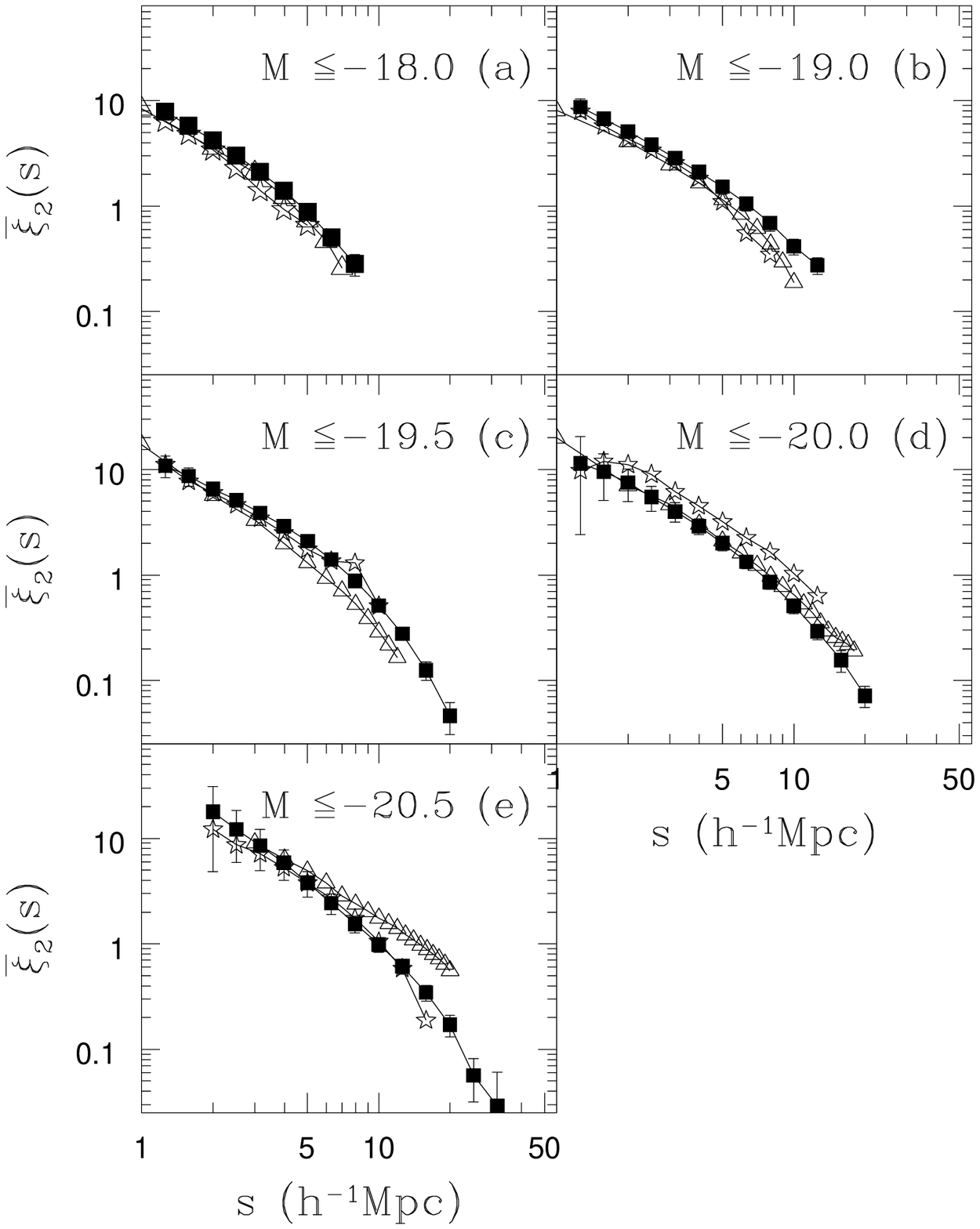]{In the five 
panels, we compare $\bar{\xi}_2(s)$ derived from the SSRS2 (squares) for each
individual sub-sample to those derived from the CfA2 south (stars) and
from the SSRS2 north (open triangles): D50 (a), D74 (b), D91 (c), D112
(d), and D138 (e).}

\figcaption[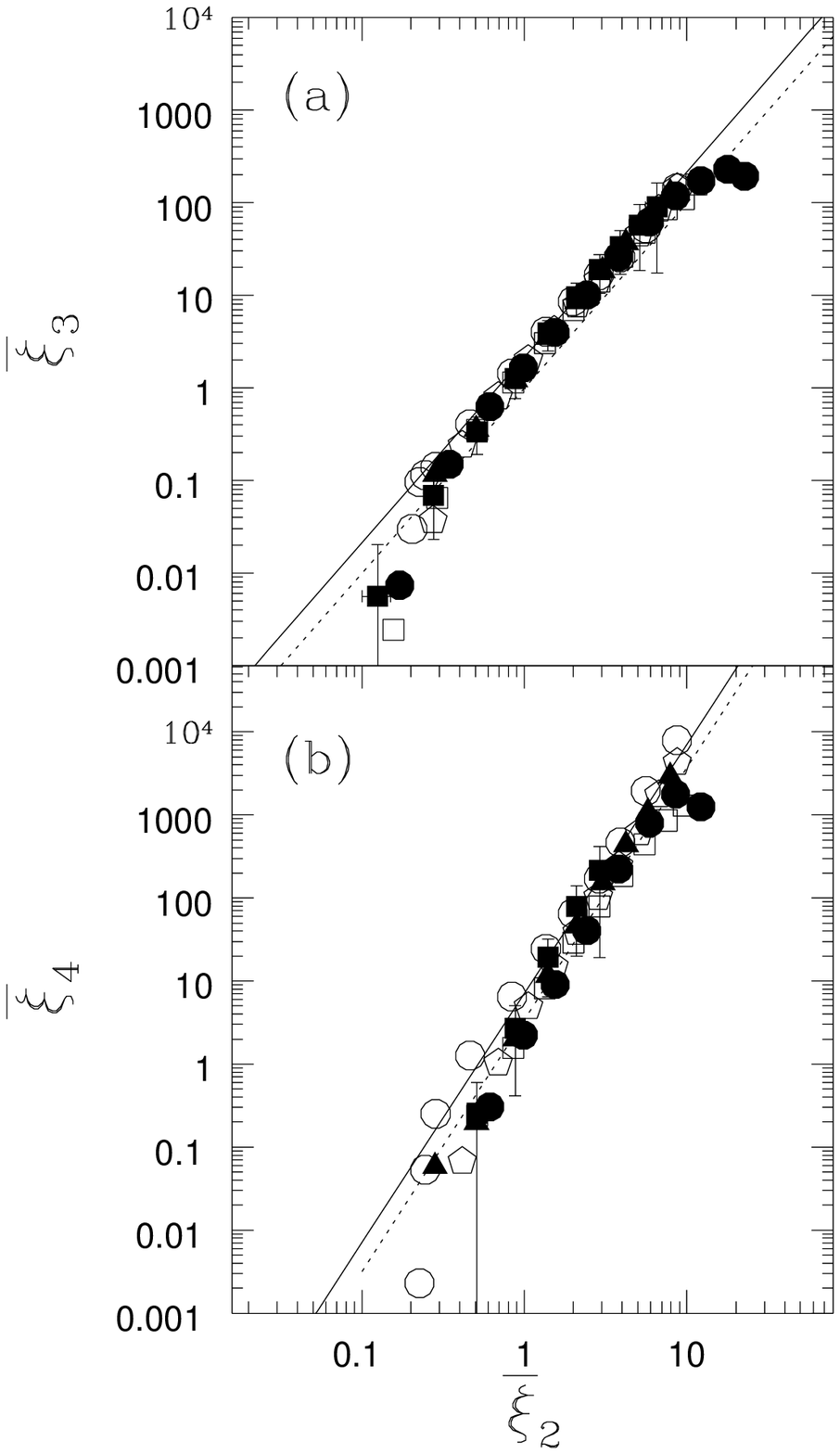]{$\bar{\xi}_2-\bar{\xi}_3$ (a) and
 $\bar{\xi}_2-\bar{\xi}_4$ (b)
for various volume-limited sub-samples identified as in Figure \ref{fig_ksi}.
The full line is the best power law fit to $L_*$ galaxies
(sub-sample D91) in the mildly non-linear regime, which corresponds
 with a high precision to the expected slope ($J-1$) of the hierarchical
models given by equation \ref{averhier}. The dotted line corresponds
 to the best fit given by Bouchet \etal 1993 with IRAS galaxies.}

\figcaption[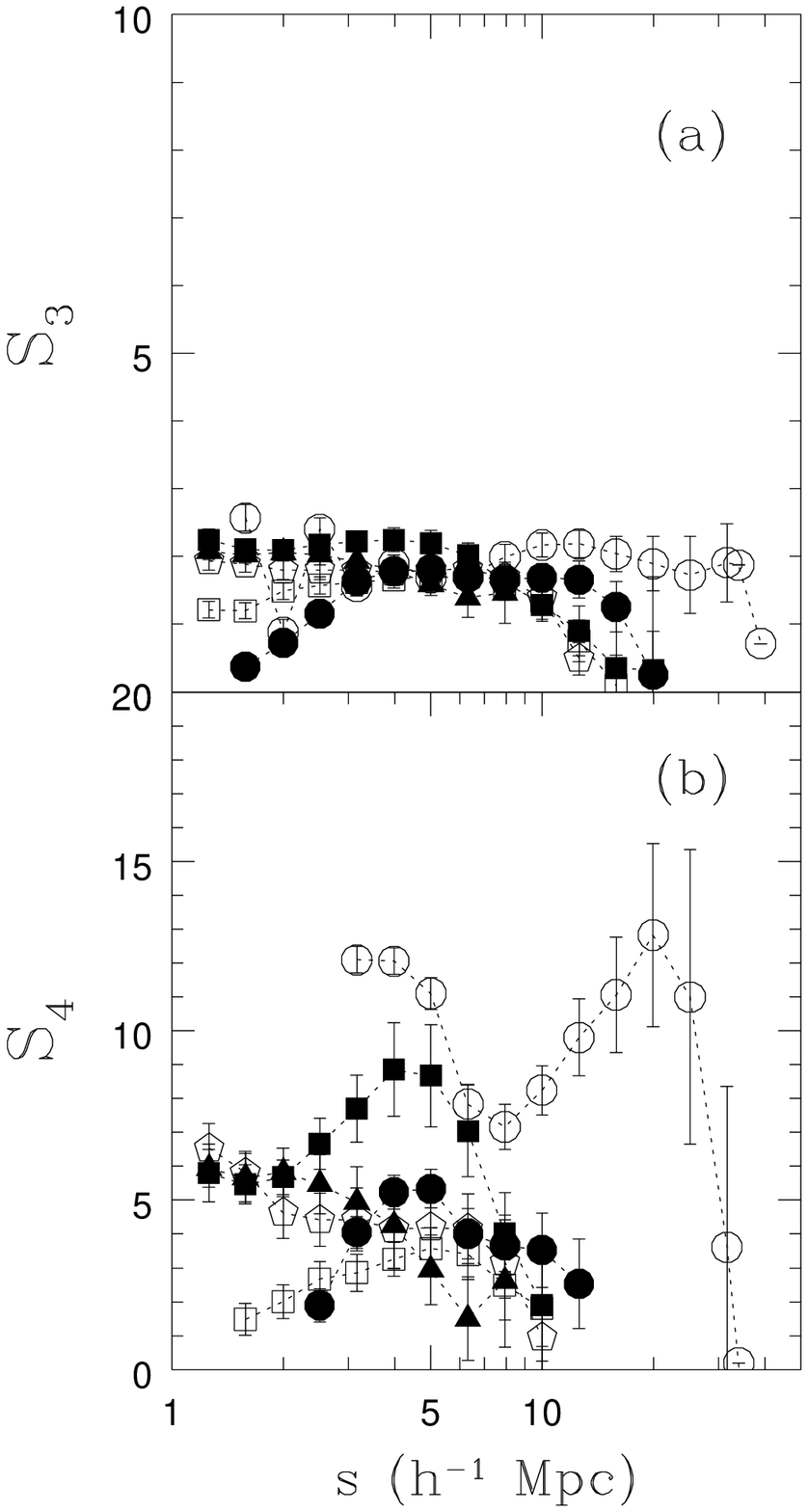]{$S_3(s)$ (a) and $S_4(s)$ (b) for various
volume-limited sub-samples identified as in Figure~\ref{fig_ksi}.}

\figcaption[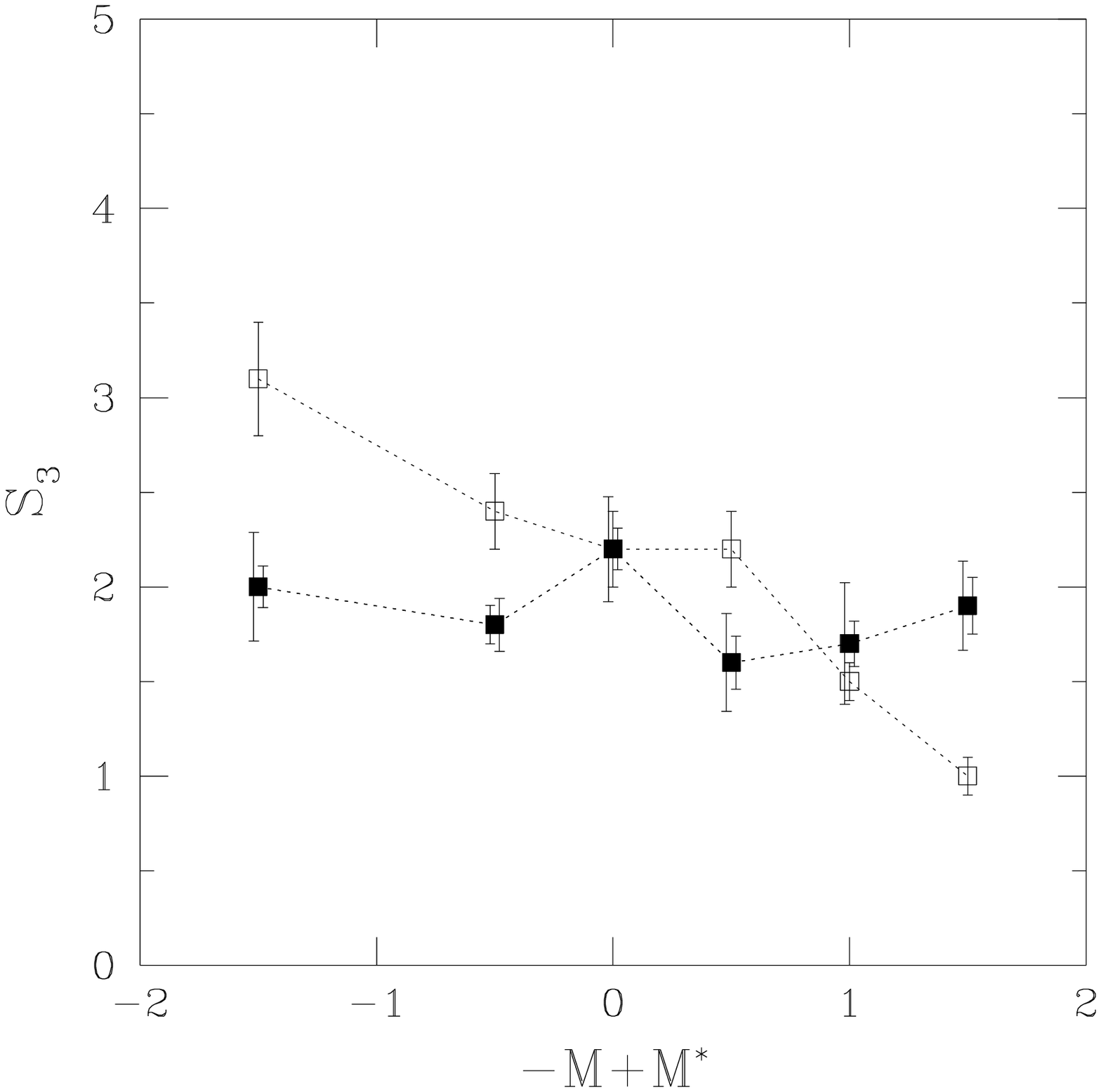]{The measured skewness of the various
  sub-samples from D50 to D168 (full squares) compared to
  $S_{3,lin}=(b_{g^*}/b_g)S_{3,g^*}$, the expected skewness in the
  context of the linear bias scenario (open squares). Two estimates of
  the errors are displayed. For sake of clarity, they are slightly
  shifted in magnitude. The left error bars are estimated from 12
  standard CDM volume-limited samples having the same geometry of the
  SSRS2, and the right error bars are estimated from the bootstrap
  method.}

\figcaption[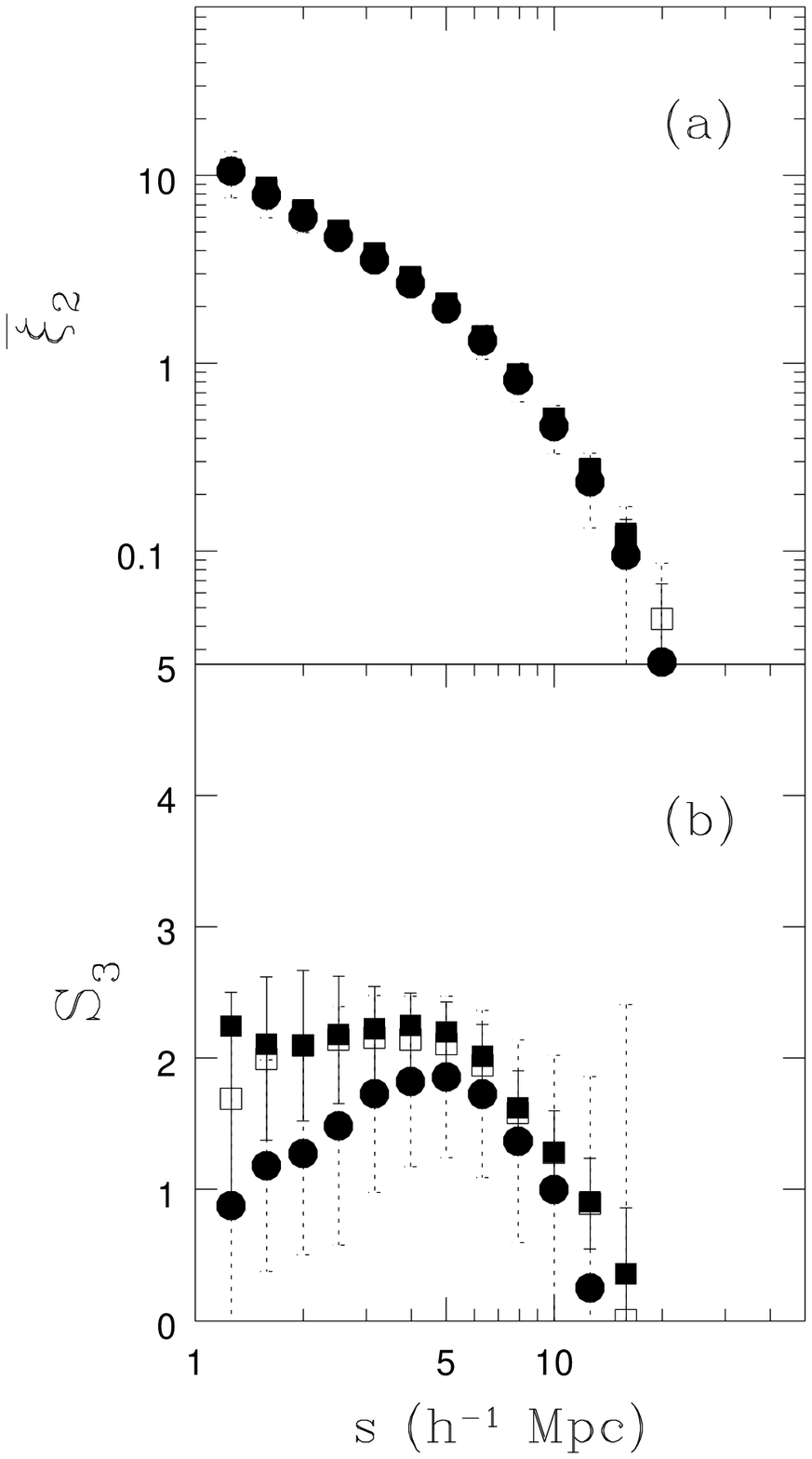]{Effect of a random dilution on 
$\bar{\xi}_2(s)$ (a) and on $S_3(s)=\bar{\xi}_3(s)/\bar{\xi}_2^2(s)$ (b).
The full squares represent the sub-sample D91, the open squares and
the full circles
represent the same sub-sample diluted respectively by a factor 2, and 
4. The error-bars are the standard 
deviation calculated over eight realizations for each dilution.}

\figcaption[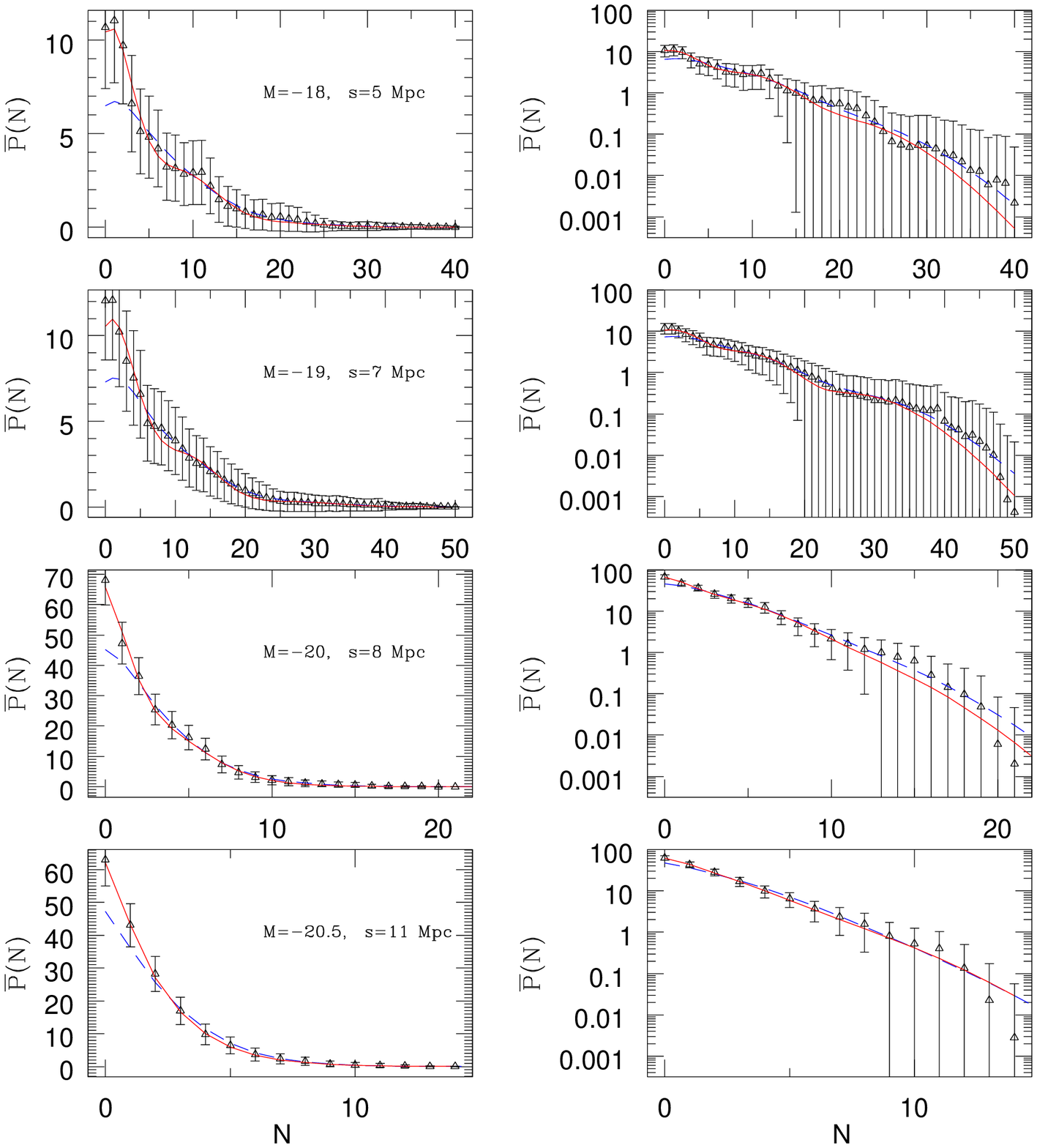]{Fits of the Edgeworth expansion
convoluted with a Poissonian to the observed number of spheres
$\bar{P(N)}$ with $N$ for 4 SSRS2 volume--limited samples ($M$
indicates the absolute magnitude limit and $s$ is the radius of the
sphere (smoothing scale).  Poissonian errors are associated to the
data (they probably represent an overestimate of the error, see Kim \&
Strauss 1998).  Dashed line: model using $\bar{\xi}_2$ and $S_3$
estimated from the moments method; solid line: best--fit using the
Edgeworth expansion at the third order.  Left column: linear scale;
right column: logarithmic scale.}

\figcaption[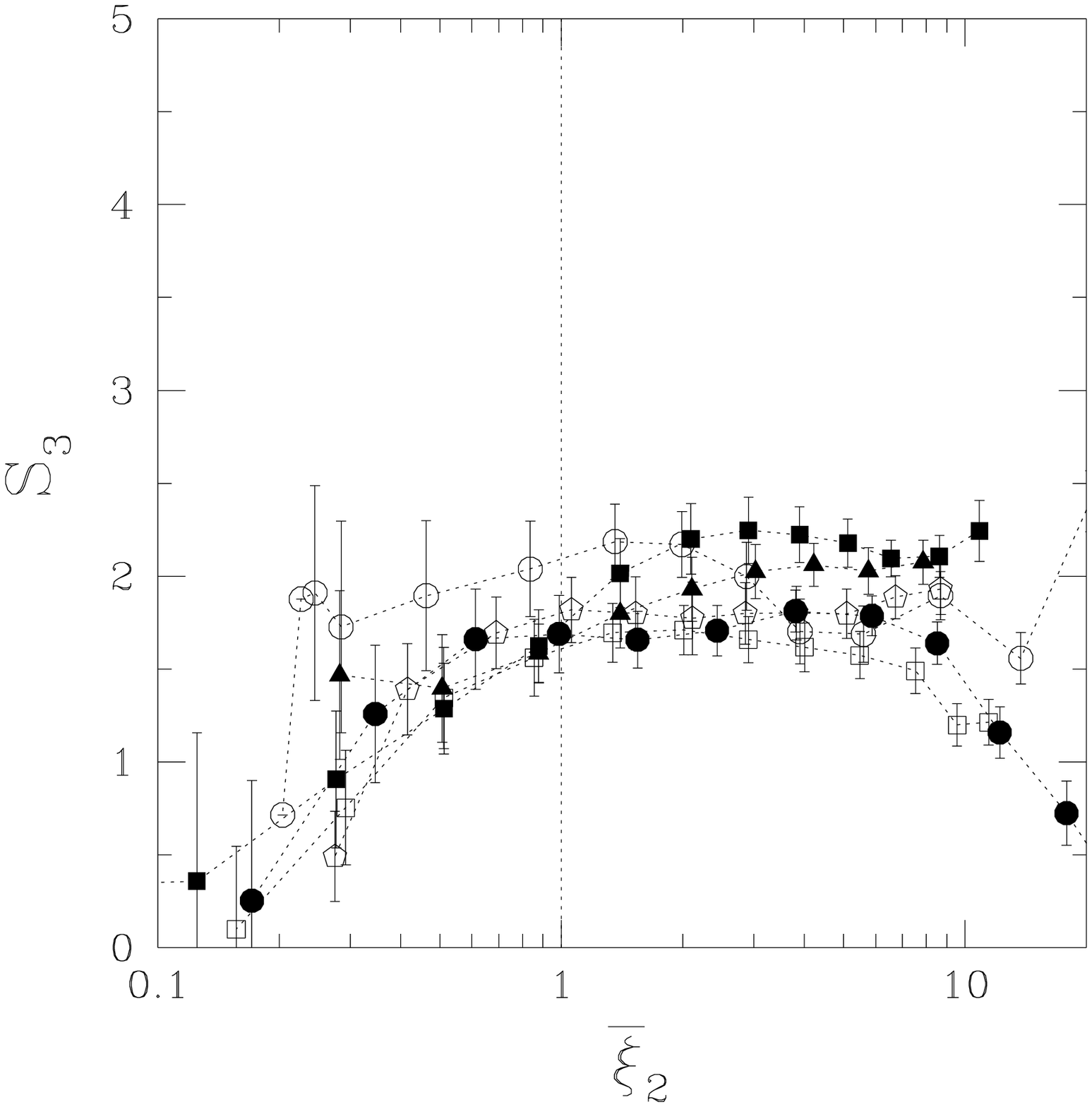]{$S_3$ as a function of $\bar{\xi}_2$ for
various sub-samples identified as in Figure~\ref{fig_ksi}.}

\figcaption[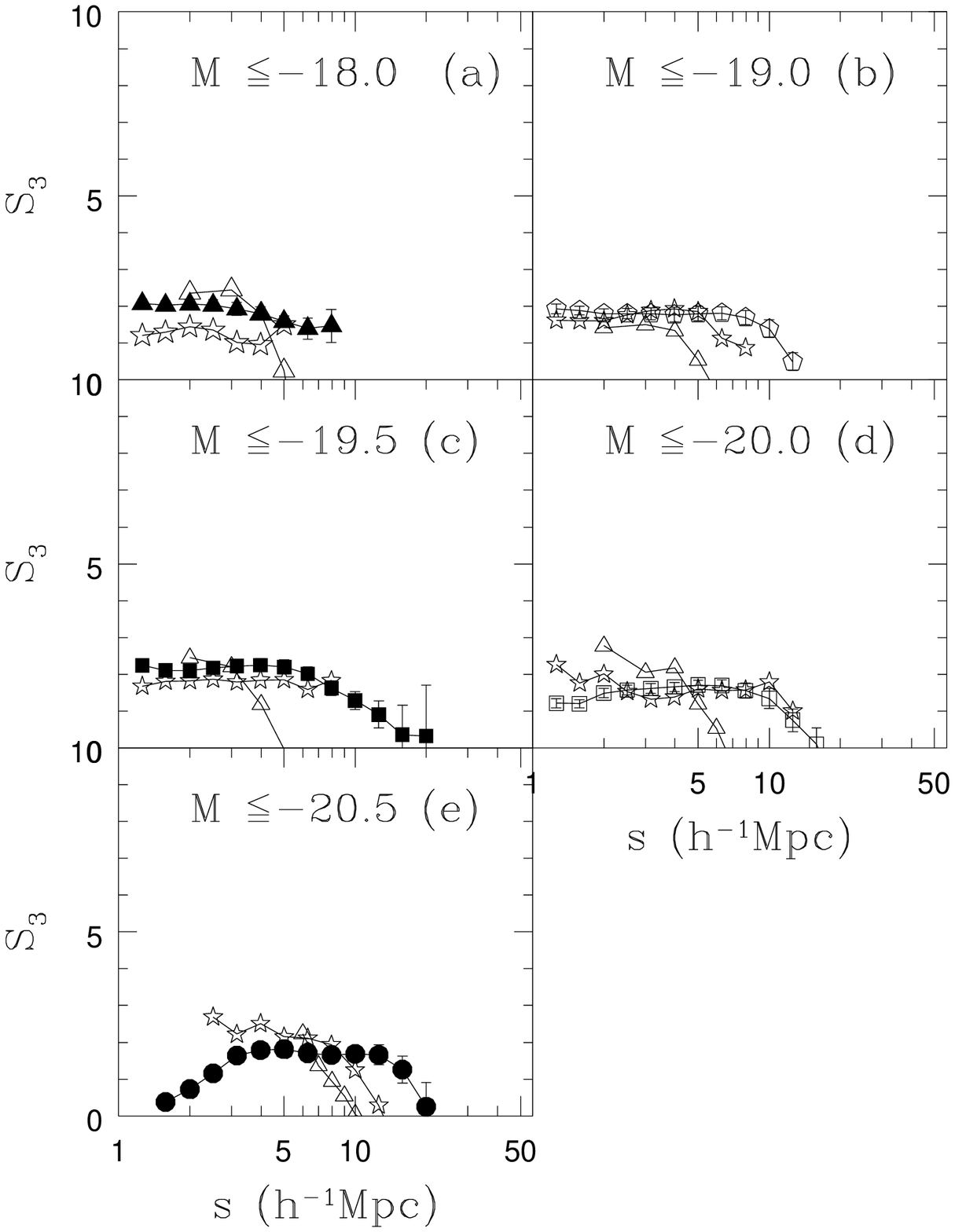]{In the five panels, we compare $S_3(s)$
derived from the SSRS2 (squares) derived from each individual
sub-sample to those derived from the CfA2 south (stars) and from the
SSRS2 north (open triangles): D50 (a), D74 (b), D91 (c), D112 (d), and
D138 (e).}

\figcaption[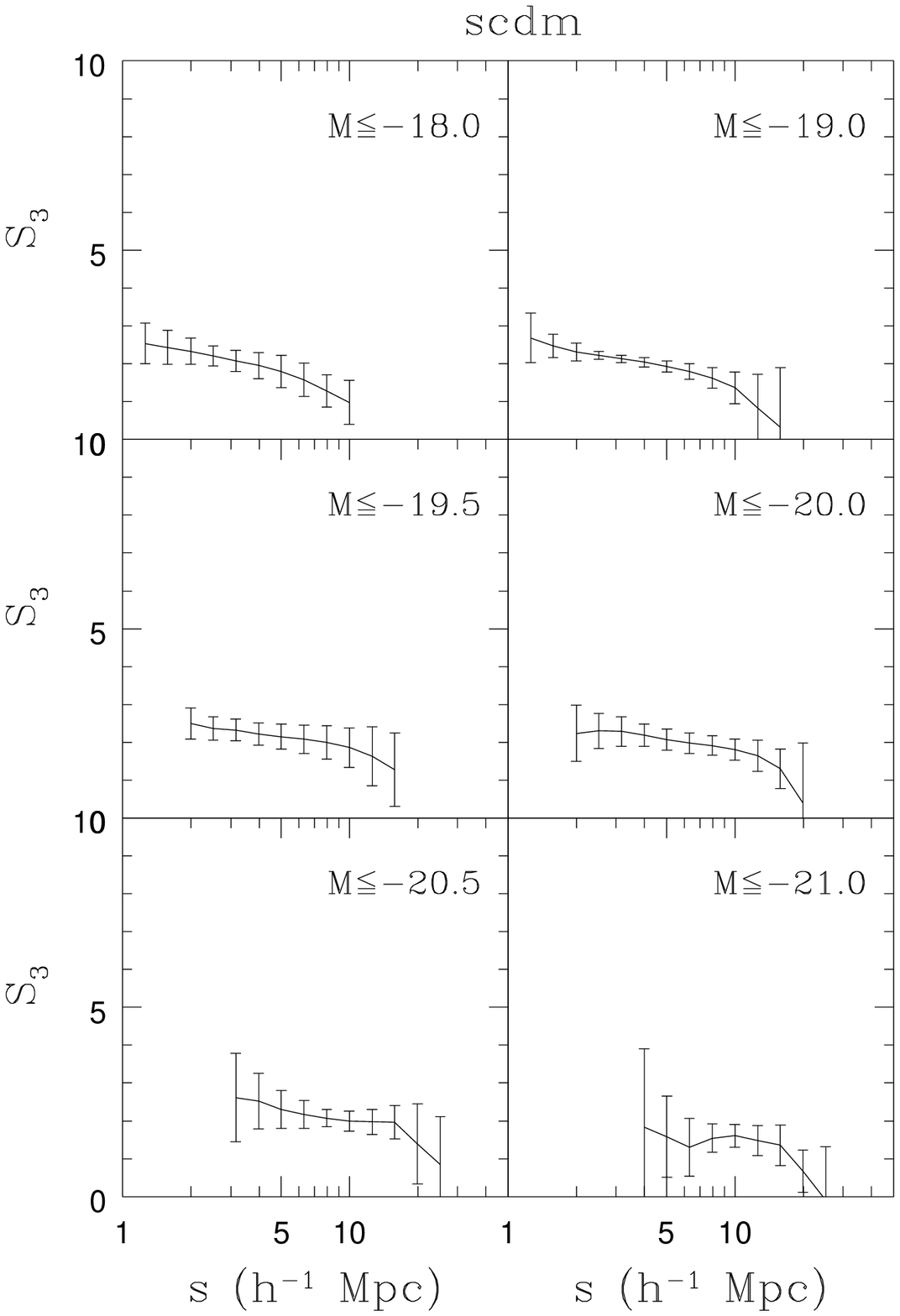]{Results of the analysis of 12 independent
standard CDM mock catalogs at the same limiting magnitudes as in
Figure \ref{fig_ksi}. The mean and the rms over these 12 samples are
shown.}

\figcaption[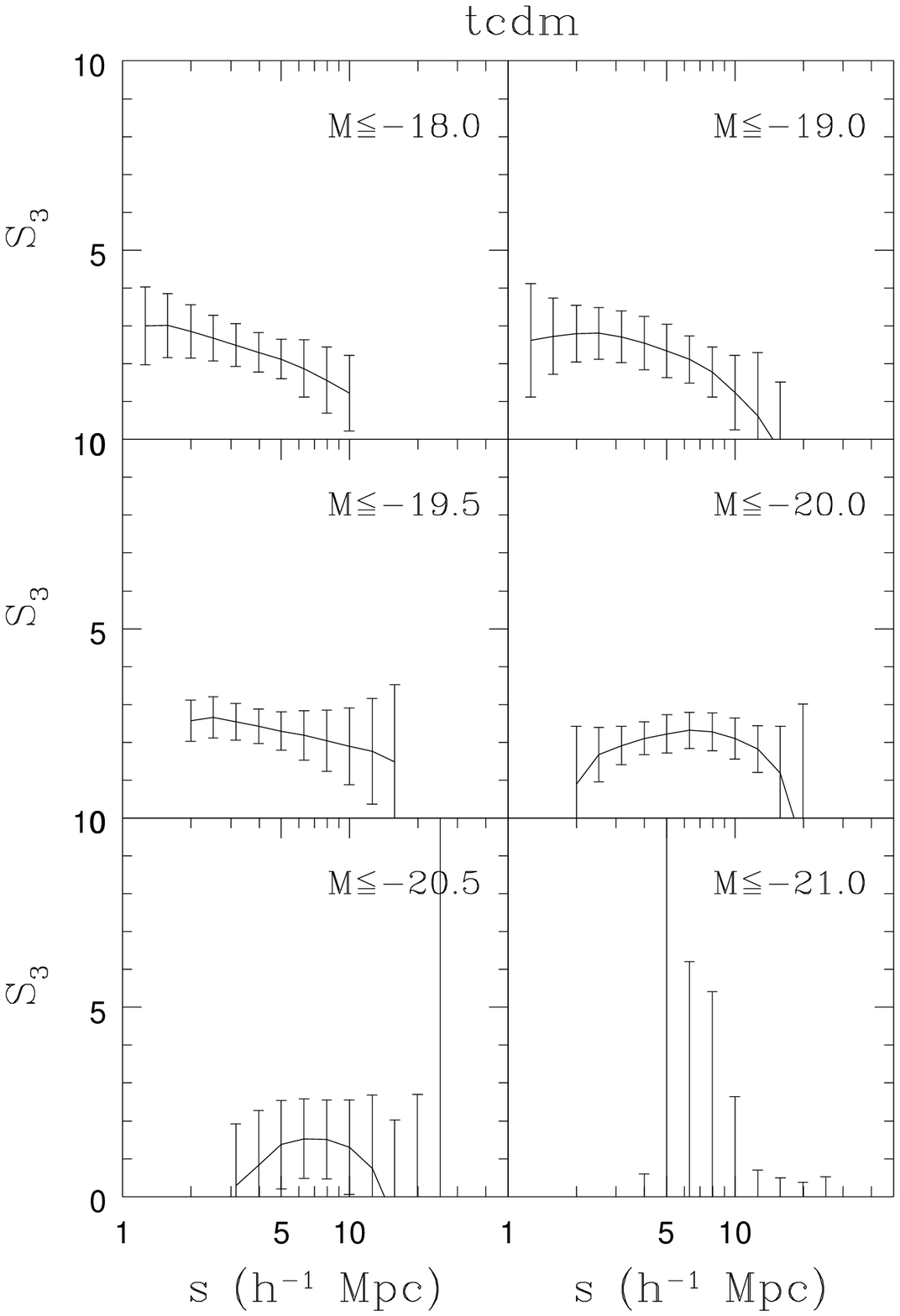]{Results of the analysis of 12 independent
tilted CDM mock catalogs at the same limiting magnitudes as in Figure
\ref{fig_ksi}. The mean and the rms over these 12 samples are shown.}

\figcaption[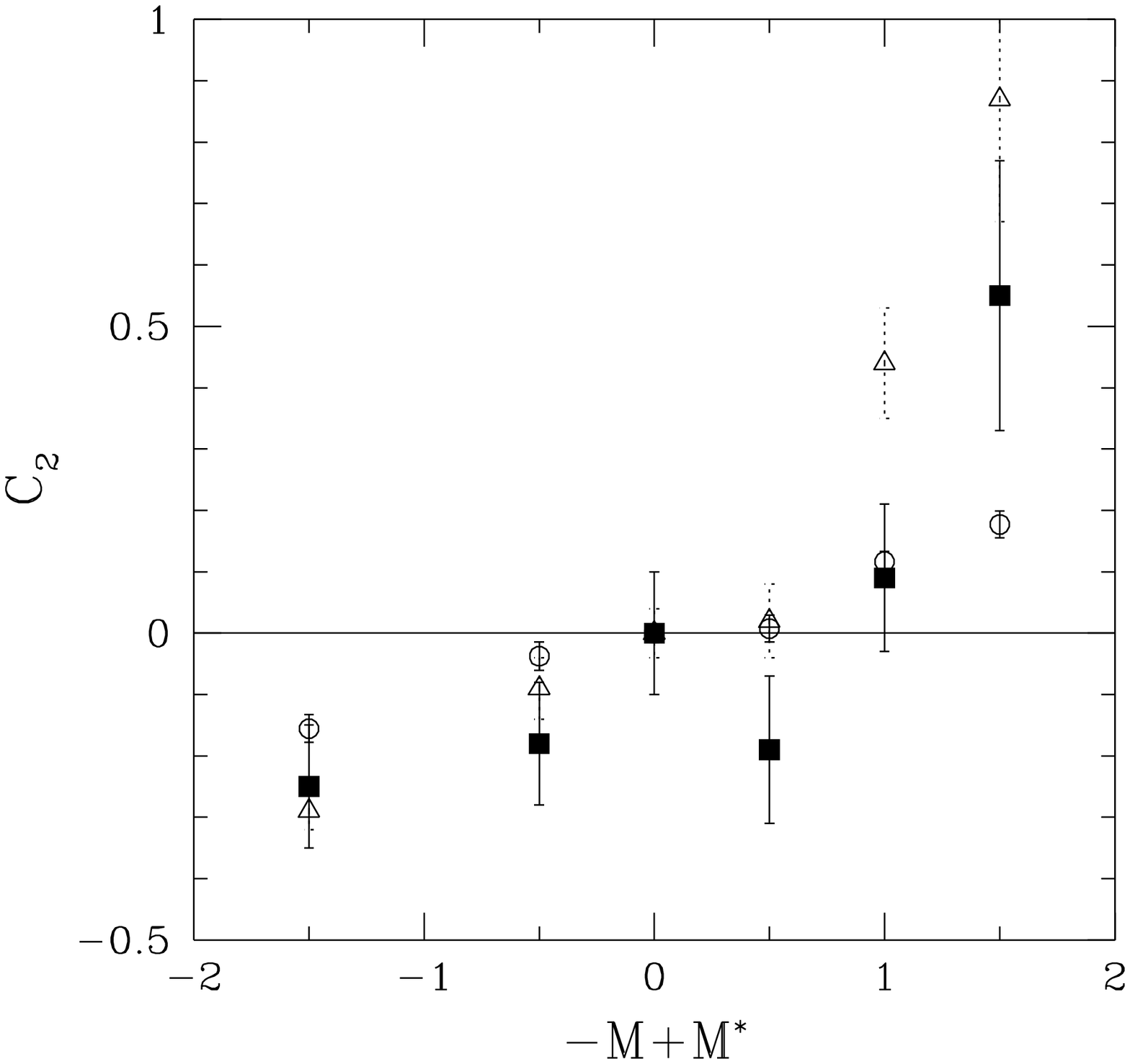]{The relative second order bias $C_2$ derived from
  the SSRS2 (full squares), and compared to the predictions of the
  model proposed by MJW in two asymptotic cases : at low redshift for
  small halos (circles), and for big halos (open squares).}

\newpage
\begin{figure} [ht]
  \plotone{pn195.ps}
 \figurenum{1}
\label{fig_pn195}
\caption{\label{fig_pn195}}
\end{figure}
\newpage
\begin{figure} [ht]
  \plotone{pn205.ps}
 \figurenum{2}
\label{fig_pn205}
\caption{\label{fig_pn205}}
\end{figure}
\newpage
\begin{figure} [ht]
  \plotone{pn21.ps}
 \figurenum{3}
\label{fig_pn21}
\caption{\label{fig_pn21}}
\end{figure}
\newpage
\begin{figure} [ht]
  \plotone{ksi2.ps}
 \figurenum{4}
\label{fig_ksi}
\caption{\label{fig_ksi}}
\end{figure}
\newpage
\begin{figure} [ht]
  \plotone{ksi2_comp.ps}
 \figurenum{5}
\label{fig_ksicomp}
\caption{\label{fig_ksicomp}}
\end{figure}
\newpage
\begin{figure} [ht]
  \plotone{ksi2324.ps}
 \figurenum{6}
\label{fig_hr}
\caption{\label{fig_hr}}
\end{figure}
\newpage
\begin{figure} [ht]
  \plotone{sn.ps}
 \figurenum{7}
\label{fig_sn}
\caption{\label{fig_sn}}
\end{figure}
\newpage
\begin{figure} [ht]
  \plotone{s3biastest.ps}
 \figurenum{8}
\label{fig_bias}
\caption{\label{fig_bias}}
\end{figure}
\newpage
\begin{figure} [ht]
  \plotone{dilu.ps}
 \figurenum{9}
\label{fig_dilu}
\caption{\label{fig_dilu}}
\end{figure}
\newpage
\begin{figure} [ht]
  \plotone{edgewor.fig.alb.ps}
 \figurenum{10}
\label{fig_edge}
\caption{\label{fig_edge}}
\end{figure}
\newpage
\begin{figure} [ht]
  \plotone{s3_ksi2.ps}
 \figurenum{11}
\label{fig_s3_ksi}
\caption{\label{fig_s3_ksi}}
\end{figure}
\newpage
\begin{figure} [ht]
  \plotone{s3_comp.ps}
 \figurenum{12}
\label{fig_s3comp}
\caption{\label{fig_s3comp}}
\end{figure}
\newpage
\begin{figure} [ht]
  \plotone{s3_scdm_mean.ps}
 \figurenum{13}
\label{fig_cdm}
\caption{\label{fig_cdm}}
\end{figure}
\newpage
\begin{figure} [ht]
  \plotone{s3_tcdm_mean.ps}
 \figurenum{14}
\label{fig_tcdm}
\caption{\label{fig_tcdm}}
\end{figure}
\newpage
\begin{figure} [ht]
  \plotone{C2.ps}
 \figurenum{15}
\label{fig_ck}
\caption{\label{fig_ck}}
\end{figure}

\end{document}